\PassOptionsToPackage{dvipsnames}{xcolor}
\documentclass[review]{elsarticle}
\usepackage{graphicx}
\usepackage{subfig}
\usepackage{ulem}
\usepackage{amsmath}
\usepackage{xcolor}
\usepackage{hyperref}
\usepackage[capitalise]{cleveref}
\usepackage[top=1.1in, bottom=1.1in, left=1.0in, right=1.0in]{geometry}
\usepackage{lineno,accsupp}
\usepackage{bm}
\usepackage{booktabs}
\usepackage{graphics}
\usepackage{multirow}
\usepackage{booktabs}
\usepackage{cases}
\usepackage{tabularx}
\usepackage{amsfonts}
\usepackage{makecell}

\makeatletter
\let\@submitted\@empty
\let\@date\@empty
\def\ps@pprintTitle{%
  \let\@oddhead\@empty
  \let\@evenhead\@empty
  \let\@oddfoot\@empty
  \let\@evenfoot\@empty
}
\makeatother

\modulolinenumbers[1]

\biboptions{numbers,sort&compress}
\begin{document}
\title{A memory-efficient deterministic method for multiscale gas flows using an ensemble-of-subproblems strategy with stochastic discrete velocities}
\author[add1]{Shuyang Zhang}
\author[add2]{Weidong Li\corref{cor1}}
\ead{lwd_1982.4.8@163.com}
\author[add2]{Ming Fang}
\author[add1,add3]{Zhaoli Guo\corref{cor1}}
\ead{zlguo@mail.hust.edu.cn}
\cortext[cor1]{Corresponding author}
  \address[add1]{State Key Laboratory of Coal Combustion, School of Energy and Power Engineering, Huazhong University of Science and Technology, Wuhan 430074, China}
  \address[add2]{National Key Laboratory of Aerospace Physics in Fluids, Mianyang 621000, China}
  \address[add3]{Institute of Interdisciplinary Research for Mathematics and Applied Science, Huazhong University of Science and Technology, Wuhan 430074, China}

\begin{abstract}
  Deterministic multiscale gas flow simulations have long suffered from the curse of dimensionality: the number of discrete velocities increases dramatically with the velocity space dimension and the Mach number, exhausting available memory and computational resources. 
  To address this issue, this paper proposes a memory-efficient deterministic method based on an ensemble-of-subproblems strategy using stochastic discrete velocities.
  This strategy transforms the originally computationally expensive problem into a series of independently and efficiently solvable subproblems.
  To be concrete, the proposed method replaces the conventional large deterministic velocity set with multiple small random velocity sets. 
  Each random set defines a subproblem, which is solved by a deterministic multiscale numerical scheme that computes macroscopic moments via Monte Carlo integration.
  The final flow field is obtained by arithmetic averaging over all subproblems.
  In this work, we employ the discrete unified gas kinetic scheme (DUGKS) for spatial discretization and term the resulting method SDV-DUGKS. 
  To validate the proposed method, several numerical test cases are conducted, including (a) the one-dimensional shock structure, (b) the two-dimensional cavity flow, and (c) supersonic flow around a square cylinder. 
  The results of the one-dimensional shock structure confirm the feasibility of the proposed method. 
  The two-dimensional cases demonstrate that, compared to its deterministic counterpart, the proposed method saves more than $80\%$ of memory usage while maintaining comparable accuracy. 
  These results indicate that the proposed method markedly reduces memory demand for multiscale flow simulations and exhibits strong potential to alleviate the curse of dimensionality that currently hinders deterministic multiscale numerical schemes from being applied to engineering problems.
\end{abstract}
\maketitle

%%\linenumbers
\section{Introduction}\label{sec1}
Multiscale gas flows exist widely in various engineering fields, such as semiconductor manufacturing~\cite{cercignani2006slow}, aerospace engineering~\cite{votta2013hypersonic}, and shale gas extraction~\cite{zhang2019coupled}. 
In such flows, characteristic length scales often vary significantly across the whole flow field, rendering the Navier–Stokes (N-S) equations based on the continuum hypothesis inadequate. 
The wide range of temporal and spatial scales involved makes accurate simulation particularly challenging, necessitating a unified modeling framework capable of describing gas dynamics across all flow regimes.
The Boltzmann equation provides a rigorous gas-kinetic-theory foundation for such a framework~\cite{grad1958principles,cercignani1988boltzmann}.

Numerical methods for solving the Boltzmann equation generally fall into two categories: stochastic and deterministic. 
The direct simulation Monte Carlo (DSMC) method~\cite{bird1994molecular,oran1998direct}, a representative stochastic approach, has long been a standard numerical tool for simulating rarefied gas flows. 
However, its grid size $\Delta{x}$ and time step $\Delta{t}$ are respectively limited by the mean free path $\lambda$ and the mean collision time $t_c$, leading to high computational costs in near-continuum and continuum regimes.
Furthermore, DSMC results are inherently affected by statistical noise~\cite{roohi2016collision}, which becomes a serious concern in low signal-to-noise-ratio scenarios.
The deterministic category, typically represented by discrete velocity methods (DVM) or discrete ordinate methods (DOM)~\cite{platkowski1988discrete,inamuro1990numerical}, can overcome this limitation.
Nevertheless, classical DVM also suffers from strict constraints on $\Delta{x}$ and $\Delta{t}$ due to the decoupling of particle transport and collision processes, making it computationally expensive when the characteristic flow scale far exceeds the kinetic scale~\cite{mieussens2014survey}. 
In recent years, with the development of asymptotically preserving (AP) multiscale schemes (e.g., the unified gas kinetic scheme (UGKS)~\cite{xu2010unified,huang2012unified}, the discrete unified gas kinetic scheme (DUGKS)~\cite{guo2013discrete,guo2015discrete,guo2021progress}, the gas kinetic Lax-Wendroff scheme (GKLWS)~\cite{Li2023a,Li2024}), the accuracy and efficiency of the discrete velocity methods for multiscale flow simulations have been significantly improved.
 
For deterministic methods, numerically solving the Boltzmann equation requires discretization in both physical space and velocity space. Regarding the velocity space discretization, the truncated velocity domain must be sufficiently large to cover the dominant part of the distribution function, and the velocity space mesh must be fine enough to minimize numerical integration errors~\cite{titarev2007conservative}.
However, an excessive number of discrete velocities leads to the curse of dimensionality~\cite{koppen2000curse}, especially in high-Mach-number and strongly non-equilibrium flow simulations.
This causes a massive increase in computational load, especially memory consumption, rapidly exhausting available computing resources, which is unacceptable for simulations in engineering problems involving complex geometries.
Therefore, it is desirable to reduce the total memory cost while maintaining satisfactory accuracy.

To mitigate the curse of dimensionality, significant research efforts have focused on reducing the number of discrete velocities used in deterministic methods. 
For instance, Mieussens proposed a conservative DVM~\cite{mieussens2000discrete}, in which the equilibrium distribution are modified to guarantee mass, momentum, and energy conservation. 
Several studies have demonstrated that the conservative DVM can achieve comparable accuracy to the conventional DVM with fewer discrete velocities~\cite{titarev2007conservative,dingwu2015study,huang2011conservative,zhang2026microscopically}. 
Although this method can reduce the number of discrete velocities, the reduction is limited.
Moreover, Kolobov~\cite{kolobov2011boltzmann} introduced a quadtree-based adaptive velocity space for homogeneous rarefied flows, which was extended by Chen~\cite{chen2012unified} to inhomogeneous flow simulations; Baranger and Brull developed a locally refined discrete velocity grid for stationary rarefied flow problems~\cite{baranger2014locally,brull2014local}. 
This adaptive discrete velocity grid technique allows the discrete velocities to be adaptively and flexibly selected according to local flow states, thereby avoiding the use of a fixed uniform velocity grid and significantly reducing the number of discrete velocities. 
However, it generally leads to complex data structures in discrete velocity space, and exchanging distribution function information at different spatial locations often requires interpolation, which may violate the conservation laws. 
In addition, Titarev~\cite{Titarev2017} and Chen~\cite{Chen2019} abandoned the Cartesian velocity grid and instead adopted unstructured grids to discretize the velocity space, in which the number of discrete velocities can be further reduced.
Nevertheless, the effects of the curse of dimensionality still persist. 
Despite these efforts, the issue of discrete velocity explosion with increasing velocity space dimensionality remains unresolved for deterministic methods.

While the aforementioned studies rely on deterministic quadrature, Monte Carlo integration is well recognized for its effectiveness in handling high-dimensional integrals~\cite{Greenbaum2012}. 
Inspired by this advantage, this work evaluates macroscopic moments via Monte Carlo integration and proposes a deterministic method using an ensemble-of-subproblems strategy with stochastic discrete velocities to alleviate the curse of dimensionality.
Multiple small stochastic discrete velocity sets are independently sampled, and each set is used to perform an independent simulation of the corresponding subproblem. 
The final result is then obtained by arithmetic averaging of the solutions over all subproblems, which reduces stochastic noise and enhances solution accuracy.
In this work, we employ the discrete unified gas kinetic scheme (DUGKS) for spatial discretization and term the resulting method SDV-DUGKS.
Since the number of discrete velocities in each set is small, the proposed method is memory efficient for simulating multiscale flow problems. 
Therefore, it holds strong potential to mitigate the curse of dimensionality.

The remainder of this paper is organized as follows. 
Section~\ref{sec2} presents the numerical algorithm of the proposed method.  
Section~\ref{sec3} validates the proposed method through several numerical test cases, with emphasis on accuracy and memory demand.
Section~\ref{sec4} concludes the paper with a concise summary.

\section{Numerical formulation and algorithm of the proposed method with stochastic discrete velocities}\label{sec2}
\subsection{The Shakhov model equation and its discretization in physical space}\label{sec21}
In this work, we consider the Shakhov model,
\begin{equation}
    \frac{\partial f}{\partial t}+\bm \xi\cdot\nabla f
    =\Omega\equiv-\frac{1}{\tau}[f-f^{S}],
\label{eq:ShakhovModel}
\end{equation}
where $ f = f(\bm x, \bm \xi, \bm \eta, \bm \zeta, t) $ is the distribution function of particles in $D$-dimensional physical space with velocity $ \bm\xi = (\xi_{1}, \dots, \xi_{D}) $ at position $ \bm x = ( x_{1}, \dots,  x_{D}) $ and time $ t $. 
Here, $ \bm\eta \in \mathbb{R}^{3-D} $ denotes the remaining velocity components in the three-dimensional velocity space; 
$\bm \zeta\in \mathbb{R}^{K}$ represents the internal degrees of freedom;
$f^S$ is the equilibrium distribution function,
\begin{equation}
    f^S=f^{eq}\Bigg[1+(1-\mathrm{Pr})\frac{\bm c\cdot \bm q}{5pRT}\Big(\frac{c^{2}+\eta^{2}}{RT}-5\Big)\Bigg],
\label{eq:ShakhovEq}
\end{equation}
with $f^{eq}={\rho}{(2\pi R T)^{-3/2}}\exp\left[-{ (c^{2}+\eta^{2}+\zeta^{2})}/{(2RT)}\right]$.
Here, $ \rho $ is the gas density, $\bm{c}=\bm{\xi} - \bm{u}$ is the peculiar velocity, $ \bm{q} $ is the heat flux, $ R $ is the gas constant, and $ T $ is the temperature. 
\iffalse
   For generality and comparability, it is essential to perform the nondimensionalization of macroscopic quantities. To accomplish this, we introduce dimensionless variables by scaling the spatial position $\bm{x}$, density $\rho$, and temperature $T$ with their respective reference values $L_{\text{ref}}$, $\rho_{\text{ref}}$, and $T_{\text{ref}}$. Correspondingly, the velocity $\bm{u}$ is nondimensionalized by the reference velocity $u_{\text{ref}} = \sqrt{2R T_{\text{ref}}}$, and time $t$ is nondimensionalized by the reference time $t_{\text{ref}} = L_{\text{ref}} / u_{\text{ref}}$. Furthermore, the heat flux $\bm{q}$ is normalized with respect to the reference heat flux $q_{\text{ref}}$, defined as $q_{\text{ref}} = \rho_{\text{ref}} R T_{\text{ref}} u_{\text{ref}}$. 
\fi
The conserved variables $ \bm W=(\rho, \rho \bm u, \rho E)^T $ are defined as the moments of the distribution function, 
\begin{equation}
    \bm W=\int \bm{\psi} f d\bm\xi d\bm\eta d\bm\zeta,
\label{eq:CMoments}
\end{equation}
with $ \bm{\psi}=\left(1,\bm\xi,\;\frac{1}{2}\left(\xi^{2}+\eta^{2}+\zeta^2\right)\right)^{T} $.  
$\rho E={\textstyle{\frac{1}{2}}}\rho u^{2}+\rho c_{V}T$ is the total energy, in which $c_{V}$ denotes the specific heat capacity at constant volume. The heat flux $\bm q$ in Eq.~\eqref{eq:ShakhovEq} is calculated from
\begin{equation}
    \bm q = \frac{1}{2}\int \bm c(c^{2}+\eta^{2}+\zeta^{2})f d\bm\xi d\bm\eta d\bm\zeta.
\label{eq:qExpression1}
\end{equation}
The relaxation time $ \tau $ is related to the dynamic viscosity $ \mu $ and pressure $ p $ by $\tau = \mu/{p}$.

To eliminate the dependence on $\bm\eta$ and $\bm\zeta$, reduced distribution functions, $g$ and $h$, are introduced to characterize the velocity field and the energy fields,
\begin{subequations}
    \begin{equation}
    g(\bm x,\bm\xi,t)=\int f d\bm\eta d\bm\zeta, \\
    \label{eq:Distributiong}
    \end{equation}
    \begin{equation}
    h(\bm x,\bm\xi,t)=\int (\eta^2+\zeta^2)fd\bm\eta d\bm\zeta.\\
    \label{eq:Distributionh}
    \end{equation}
    \label{eq:Distributionf}%
\end{subequations}
Correspondingly, the reduced equilibrium distribution functions, $g^S$ and $h^S$, are given by
\begin{subequations}
    \begin{equation}
    \begin{aligned}
    \quad g^S={g^{eq}}\left\{1+(1-\mathrm{Pr}){\frac{\bm c\cdot \bm q}{5p R T}}\left[\frac{c^{2}}{R T}-D-2\right]\right\},
    \label{eq:ReducedgS}
    \end{aligned}
    \end{equation}
    \begin{equation}
    \begin{aligned}
    h^S
    =(3-D+K)RT{g}^{eq}+(1-\mathrm{Pr}){\frac{\bm c\cdot \bm q}{5p R T}}\left[\left(\frac{c^{2}}{RT}-D\right)(3-D+K)-2K\right]RT{g}^{eq},
    \label{eq:ReducedhS}
    \end{aligned}
    \end{equation}
    \label{eq:ReducedfS}%
\end{subequations}
with ${g^{eq}}={\rho}{(2\pi R T)}^{-D/2}\exp{\left[-{c^{2}}/{(2RT)}\right]}$.

For conciseness, the kinetic equations for $g$ and $h$ can be simplified as
\begin{equation}
    {\frac{\partial \phi}{\partial t}}+\bm\xi\cdot\nabla \phi=\Omega\equiv-{\frac{1}{\tau}}[\phi-\phi^{S}], \\
\label{eq:phiEquation}
\end{equation}
where $ \phi=g \ \text{or} \ h $, $ \phi^S=g^S \ \text{or} \ h^S $.
Eq.~\eqref{eq:phiEquation} can be discretized in physical space by any kind of multiscale scheme. 
In this work, the DUGKS~\cite{guo2013discrete,guo2015discrete,guo2021progress} is adopted.
Integrating Eq.~\eqref{eq:phiEquation} over the control volume $ V_j $ centered at $ \bm x_j $ from $ t_n $ to $ t_{n+1} $ yields the evolution equation for $ \phi_{j} $,
\begin{equation}
    \phi_{j}^{n+1}-\phi_{j}^{n}+\frac{\Delta t}{|V_{j}|}\bm F^{n+1/2}
    =\frac{\Delta t}{2}\big[\Omega_{j}^{n+1}+\Omega_{j}^{n}\big], \\
\label{eq:phievelotion}
\end{equation}
where the convection term and collision term are integrated based on the midpoint rule and trapezoidal rule, respectively. The term $ \bm{F}^{n+1/2} $, defined as the microflux across the cell interface, is given by
\begin{equation}
    \bm F^{n+1/2}(\bm\xi)=\int_{\partial V_{j}}\,(\bm\xi\cdot \bm n)\phi(\bm{x},\bm\xi,t_{n+1/2})\,d \bm S, \\
\label{eq:microfluxF}
\end{equation}
where $\partial V_{j}$ denotes the surface of the cell, and $\bm n$ is the outward unit normal vector to the surface $\partial V_{j}$.

To remove the implicitness of the collision term, DUGKS introduces the auxiliary distributions,
\begin{equation}
    \tilde{\phi}=\phi-{\frac{\Delta t}{2}}\Omega,\quad \tilde{\phi}^{+}=\phi+{\frac{\Delta t}{2}}\Omega. \\
\label{eq:auxiliaryphitilde}
\end{equation}
Eq.~\eqref{eq:phievelotion} can be further rewritten as
\begin{equation}
    \tilde{\phi}_{j}^{n+1}=\tilde{\phi}_{j}^{+,n}-\frac{\Delta t}{\mid V_{j}\mid}\bm F^{n+1/2}. \\
\label{eq:tildephievelution}
\end{equation}
Based on the conservation properties of the collision term, the conserved variables are computed by
\begin{equation}
    \rho=\int\tilde{g}d\bm\xi,\quad\rho \bm u=\int\bm\xi\tilde{g}d\bm\xi,
    \quad\rho E=\frac{1}{2}\int(\xi^{2}\tilde{g}+\tilde{h})d\bm\xi, \\
\label{eq:gTlideCMoments}
\end{equation}
and heat flux $\bm q$ is calculated from
\begin{equation}
    \bm q=\frac{2\tau}{2\tau+\Delta t\mathrm{Pr}}\tilde{\bm q},\,\quad \mathrm{with}\ \quad
    \tilde{\bm q}=\frac{1}{2}\int\bm c(c^{2}\tilde{g}+\tilde{h})\mathrm{d}\bm{\xi}. \\
\label{eq:gTlideq}
\end{equation}
Given the microflux $ \bm F^{n+1/2} $, the distribution function $\tilde{\phi}$ can be explicitly updated via Eq.~\eqref{eq:tildephievelution}.

To compute $\bm F^{n+1/2}$, the distribution function $\phi^{n+1/2}$ at the cell interface center is obtained by the characteristic line solution of Eq.~\eqref{eq:phiEquation}. That is,
\begin{equation}
\begin{aligned}
    \phi\left(\bm x_{b},\bm\xi,t_{n}+s\right)-\phi\left(\bm x_{b}-\bm\xi s,\bm\xi,t_{n}\right)=\frac{s}{2}\left[\Omega\left(\bm x_{b},\bm\xi,t_{n}+s\right)
    +\Omega\left(\bm x_{b}-\bm\xi s,\bm\xi,t_{n}\right)\right],\\
\end{aligned}
\label{eq:halfphievelotion}
\end{equation}
where $s=\Delta{t}/2$, $\bm x_b$ is the interface center of cell $j$. 
Defining $\bar{\phi}$ and $\bar{\phi}^{+}$ as
\begin{equation}
    \bar{\phi}=\phi-{\frac{s}{2}}\Omega,\quad \bar{\phi}^{+}=\phi+{\frac{s}{2}}\Omega, \\
\label{eq:auxiliaryphibar}
\end{equation}
one can further simplify Eq.~\eqref{eq:halfphievelotion} as
\begin{equation}
    \bar{\phi}(\bm x_{b},\bm\xi,t_{n+1/2})=\bar{\phi}^{+}(\bm x_{b}-\bm\xi s,\bm\xi,t_{n}), \\
\label{eq:barphievelotion}
\end{equation}
where $ \bar{\phi}^{+}(\bm x_{b}-\bm\xi s,\bm\xi,t_{n}) $ is obtained via linear reconstruction,
\begin{equation}
    \bar{\phi}^{+}(\bm x_{b}-\bm\xi s,\bm\xi,t_{n})=\bar{\phi}^{+}(\bm x_{j},\bm\xi,t_{n})
    +(\bm x_{b}-\bm x_{j}-\bm\xi s)\cdot\bm\sigma_{j}, \\
\label{eq:phiBarinterfaceGot}
\end{equation}
with $\bm\sigma_{j}$ denoting the slope of $\bar{\phi}^{+}$ in cell $j$. 
Upon obtaining $\bar{\phi}(\bm x_{b},\bm\xi,t_{n+1/2})$, the macroscopic variables at time $t_{n+1/2}$ can be computed by
\begin{equation}
    \rho=\int\bar{g}d\bm\xi,\quad\rho \bm u=\int\bm\xi\bar{g}d\bm\xi,
    \quad\rho E=\frac{1}{2}\int(\xi^{2}\bar{g}+\bar{h})d\bm\xi, \\
\label{eq:gBarCMoments}
\end{equation}
and
\begin{equation}
    \bm q=\frac{2\tau}{2\tau+s\mathrm{Pr}}\bar{\bm q},\,\quad \mathrm{with}\ \quad
    \bar {\bm q}=\frac{1}{2}\int\bm c(c^{2}\bar{g}+\bar{h})\mathrm{d}\bm{\xi}. \\
\label{eq:gBarq}
\end{equation}
Thus, the Shakhov equilibrium distribution function $\phi^{S}(\bm x_{b},\bm\xi,t_{n}+s)$ can be obtained from Eq.~\eqref{eq:ReducedfS}.
Finally, the original distribution function at the cell interface is then calculated via Eq.~\eqref{eq:auxiliaryphibar},
\begin{equation}
    \phi\bigl(\bm x_b,\bm\xi,t_{n+1/2}\bigr)=\frac{2\tau}{2\tau+s}\bar{\phi}\bigl(\bm x_b,\bm\xi,t_n+s\bigr)
    +\frac{s}{2\tau+s}\phi^S\bigl(\bm x_b,\bm\xi,t_n+s\bigr). \\
\label{eq:phiBarShakhov}
\end{equation}

For brevity, only some essential aspects of the implementation of the DUGKS are described in the present section, and more numerical details can be found elsewhere~\cite{guo2015discrete}.

\subsection{Discretization of particle velocity space via Monte Carlo integration points}\label{sec22}
In Sec.~\ref{sec21}, the discretization of the Shakhov model equation in physical space was discussed. 
The discretization of particle velocity space is jointly determined by the required moment integral constraints and deterministic quadrature rules, such as half-range Hermite quadrature~\cite{Gautschi2016}, Hermite quadrature~\cite{Yang1995}, Newton-Cotes quadrature~\cite{hayes1970proof}. 
With the velocity space discretized into $N_d$ discrete velocity points, Eq.~\eqref{eq:phiEquation} becomes
\begin{equation}\label{eq:DVphiEquation}
    {\frac{\partial \phi_\alpha}{\partial t}}+\bm\xi_\alpha\cdot\nabla \phi_\alpha=\Omega\equiv-{\frac{1}{\tau}}[\phi_\alpha-\phi_\alpha^{S}], \qquad \alpha = 1, 2, \cdots, N_d,
\end{equation}
where $\bm\xi_\alpha$ is the $\alpha$-th discrete velocity point, and $\phi_\alpha = \phi(\bm x, \bm \xi_\alpha, t)$ is the corresponding distribution function.
Accordingly, the macroscopic variables are computed by
\begin{equation}\label{eq:sum_macro_moments_deter}
    \begin{pmatrix}\rho
        \\\rho\bm{u}
        \\2\rho{E}
        \\2\bm{q}
    \end{pmatrix}=\sum_\alpha^{N_d}\omega_{\alpha}[\begin{pmatrix}1
                           \\\bm\xi_{\alpha}
                           \\\xi_{\alpha}^2
                           \\\bm{c}_{\alpha}(c_{\alpha})^2
                           \end{pmatrix}\phi_{g,\alpha}
                           +\begin{pmatrix}0
                           \\\bm{0}
                           \\1
                           \\\bm{c}_{\alpha}
                           \end{pmatrix}\phi_{h,\alpha}],
\end{equation}
where $\omega_{\alpha}$ denotes the quadrature weight associated with the discrete velocity point $\bm{\xi}_\alpha$, $\phi_{g,\alpha} = g_\alpha$, $\phi_{h,\alpha} = h_\alpha$.

As noted in the introduction, deterministic quadrature rules for the velocity space discretization leads to computational costs that grow exponentially with the velocity space dimensionality $D$. 
To address this problem, we replace the deterministic quadrature with Monte Carlo integration~\cite{robert1999monte}.
Specifically, the Monte Carlo integration procedure consists of two core components: sampling and moment evaluation, as detailed in the following subsections.

\subsubsection{Monte Carlo integration in velocity space}\label{sec221}
We begin with the moment evaluation.
Using Monte Carlo integration, $N_r$ velocity points are randomly sampled from the velocity space and Eq.~\eqref{eq:sum_macro_moments_deter} is reformulated as
\begin{equation}\label{eq:sum_macro_moments_stoca}
    \begin{pmatrix}\rho
 \\\rho\bm{u}
 \\2\rho{E}
 \\2\bm{q}
\end{pmatrix}=\frac{V_{pvs}}{N_r}\sum^{N_r}_{\alpha=1}[\begin{pmatrix}1
                           \\\bm\xi_\alpha
                           \\\xi_\alpha^2
                           \\\bm{c}_\alpha(c_\alpha)^2
                           \end{pmatrix}g_{\alpha}
                           +\begin{pmatrix}0
                           \\\bm{0}
                           \\1
                           \\\bm{c}_\alpha
                           \end{pmatrix}h_{\alpha}],
\end{equation}
where $V_{pvs}$ denotes the volume of the truncated velocity space domain utilized in simulations. 
Notably, Monte Carlo integration provides unbiased estimators of the expected value of a distribution, and its error convergence rate, $O(N_r^{-1/2})$, is independent of dimensionality~\cite{caflisch1998Monte,owen1992central}. 
Naturally, increasing the sample size $N_r$ improves the Monte Carlo integration accuracy. 
To avoid increasing memory requirements, an ensemble-of-subproblems strategy is adopted, where $M$ independent Monte Carlo integrations are performed, and the final quadrature result is obtained through arithmetic averaging over $M$ realizations.
This averaging achieves an error convergence rate of $O((MN_r)^{-1/2})$ without increasing the per-realization memory usage.

In the proposed method, we conduct $M$ independent simulations. 
In each simulation, a set of $N_r$ discrete velocity points is independently sampled, which is used for moment evaluations and to evolve the solution following the DUGKS algorithm until convergence.
The final simulation result is obtained by averaging the $M$ converged solutions, i.e.,
\begin{equation}
    \begin{pmatrix}\rho
 \\\rho\bm{u}
 \\\rho{E}
 \\\bm{q}
\end{pmatrix}=\frac{1}{M}\sum_{m=1}^{M}\begin{pmatrix}\rho_{(m)}
 \\\rho_{(m)}\bm{u}_{(m)}
 \\\rho_{(m)}{E}_{(m)}
 \\\bm{q}_{(m)}
\end{pmatrix}, \\
\label{eq:average_M}
\end{equation}
in which the subscript $m$ denotes the $m$-th realization and $M$ is the total number of realizations. 
For the $m$-th realization, the macroscopic variables $(\rho_{(m)}, \rho_{(m)}\bm{u}_{(m)}, \rho_{(m)}{E}_{(m)}, \bm{q}_{(m)})^{T}$ are computed using Eq.~\eqref{eq:sum_macro_moments_stoca} with the independently sampled discrete velocity set $\{\boldsymbol{\xi}_{\alpha,(m)}\} = \{\boldsymbol{\xi}_{1,(m)}, \boldsymbol{\xi}_{2,(m)}, \cdots, \boldsymbol{\xi}_{N_r,(m)}\}$.
If $N_r$ is smaller than $N_d$, the proposed method is more memory-efficient than conventional deterministic methods.

\subsubsection{Sampling of stochastic discrete velocities in particle velocity space}\label{sec222}
\begin{figure}[!ht]
\centering
  \subfloat{\includegraphics[width=0.27\textwidth]{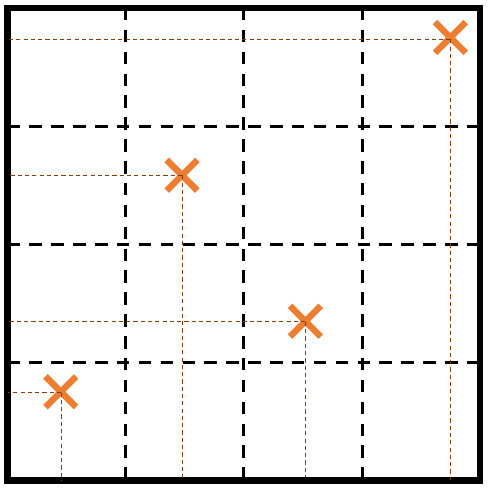}}~~
 \caption{An example distribution of $4$ samples generated by LHS in a two-dimensional parameter space.}
\label{LHS_221}
\end{figure}

For Monte Carlo integration, a good sampling technique is crucial for both accuracy and efficiency.
In this work, the Latin hypercube sampling (LHS)~\cite{loh1996latin,mckay2000comparison} is employed to sample velocity points.
LHS combines stratified sampling with random pairing. 
It ensures that each dimension is stratified with exactly one sample projection per stratum, as illustrated with $4$ samples in $2\mathrm{D}$ parameter space in Fig.~\ref{LHS_221}, thereby effectively preventing the clustering of sample points. 
Consequently, compared with simple random sampling, LHS enhances the coverage of the parameter space, making it an efficient sampling technique for Monte Carlo integration~\cite{mckay2000comparison}. 

\begin{figure}[!ht]
\centering
  \subfloat{\includegraphics[width=0.9\textwidth]{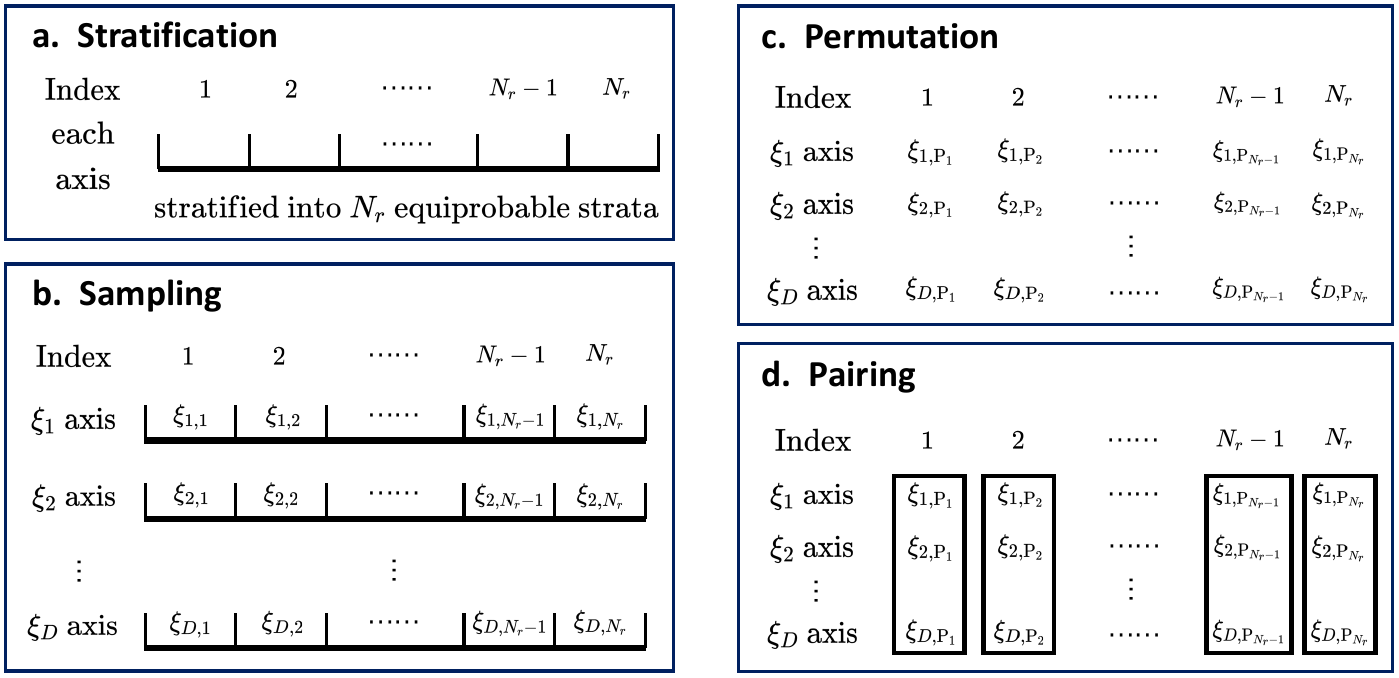}}~~
 \caption{The schematic of LHS implementation in a three-dimensional parameter space with $N$ samples.}
\label{LHS_222}
\end{figure}

Considering the task of sampling $N_r$ velocity points from a truncated velocity space $[-a,a]^D $ ($a > 0$), the LHS procedure consists of four steps: stratification, sampling, permutation, and pairing, as illustrated in Fig.~\ref{LHS_222}. 
These steps are detailed as follows:
\begin{enumerate}[(a)]
  \item Stratification: \\
    Stratify each dimension of the truncated velocity space $[-a,a]^D $ into $N_r$ disjoint subintervals of equal length $\Delta v=2a/N_r$:
    \begin{equation}
       [-a, -a+\Delta v), \ [-a+\Delta v, -a+2\Delta v), \ \cdots .
    \end{equation}
  \item Sampling:      \\
    Randomly sample one value within each stratum, yielding ordered sample sets for each dimension:
    \begin{equation}
       \{\xi_{1,1},\cdots,\xi_{1,{N_r}}\}, \ \{\xi_{2,1},\cdots,\xi_{2,{N_r}}\}, \ \cdots, \ \{\xi_{D,1},\cdots,\xi_{D,{N_r}}\}. 
    \end{equation}
  \item Permutation:     \\
    For each dimension, independently and randomly permute the order of the drawn values to generate the sequences:
    \begin{equation}
       \{\xi_{k,{\mathrm{P}_1}},\xi_{k,{\mathrm{P}_2}},\cdots,\xi_{k,{\mathrm{P}_{N_r}}}\}, \quad k=1,\cdots, D,
    \end{equation}
    where $\mathrm{P}_\alpha$ denotes the index of the $\alpha$-th element after permutation.
  \item Pairing:          \\
    Form $N_r$ velocity points by taking the $\alpha$-th element from each permuted sample set to construct the $\alpha$-th velocity point:
    \begin{equation}
      \bm\xi_\alpha = (\xi_{1,{\mathrm{P}_\alpha}},\xi_{2,{\mathrm{P}_\alpha}}, \cdots, \xi_{D,{\mathrm{P}_\alpha}}) \in [-a,a]^D, \quad \alpha = 1,\cdots, N_r. 
    \end{equation}
\end{enumerate}
In each realization, $N_r$ velocity points are generated by the above LHS procedures, and subsequently used in the simulation.

\subsection{The microscopically conservation-enforced correction}\label{sec23}
Theoretically, the collision terms $\Omega_g$ and $\Omega_h$ obey the conservative constraints, i.e.,
\begin{subequations}
    \begin{equation}
    \int\Omega_g{d}\bm\xi=0, \\
    \label{eq:conservation_mass}
    \end{equation}
    \begin{equation}
    \int\bm\xi\Omega_g{d}\bm\xi=0, \\
    \label{eq:conservation_momentum}
    \end{equation}
    \begin{equation}
    \int(\xi^2\Omega_g+\Omega_h){d}\bm\xi=0. \\
    \label{eq:conservation_energy}
    \end{equation}
    \label{eq:conservation_laws}%
\end{subequations}
For numerical accuracy and physical consistency, this condition must be enforced in numerical simulations.
However, when these integrals are approximated by numerical integration, the conservation property of the collision terms is not satisfied exactly.
The violation of conservation constraints gives rise to the unphysical numerical source term $\left({1}/{\tau}\right)\bm{R}$, with $\bm{R}$ in this work given by
\begin{equation}
    \bm{R}=\tau\frac{V_{pvs}}{N_r}\sum^{N_r}_{\alpha=1}(\begin{pmatrix}1
                           \\\bm\xi_\alpha
                           \\{\xi_\alpha}^2
                           \end{pmatrix}\Omega_{g,\alpha}
                           +\begin{pmatrix}0
                           \\\bm{0}
                           \\1
                           \end{pmatrix}\Omega_{h,\alpha})
          =\frac{V_{pvs}}{N_r}\sum^{N_r}_{\alpha=1}\begin{pmatrix}{g}^S_\alpha
                           \\\bm\xi_{\alpha}{g}^S_\alpha
                           \\(\xi_{\alpha}^2{g}^S_\alpha+{h}^S_\alpha)
                           \end{pmatrix}
                          -\begin{pmatrix}\rho
                          \\\rho\bm{u}
                          \\2\rho{E}
                          \end{pmatrix}, \\
\label{eq:numerical_error}
\end{equation}
which accumulates over successive iterations and may adversely affect numerical stability~\cite{titarev2007conservative,zhang2026microscopically}.

To reduce the unphysical term $\bm{R}$ and enhance numerical stability, the microscopically conservation-enforced correction~\cite{mieussens2000discrete,titarev2007conservative} is adopted in the present study.
The proposed method incorporates the correction steps within the original DUGKS, denoted as MicroC-DUGKS~\cite{zhang2026microscopically}.
These steps specifically revise the Shakhov equilibrium distribution to ensure the conservation of mass, momentum, and total energy, thereby directly eliminating the unphysical term $\bm{R}$.
For completeness and clarity, a schematic of the correction implementation is provided below (full details are available in Refs.~\cite{titarev2007conservative,zhang2026microscopically}).

To make the unphysical term $\bm{R}$ vanish, a variable $\bm{A}^* = (\rho^*, \bm{u}^*, T^*, \bm{q}^*)^T$ (corrected state of $\bm{A}=(\rho, \bm{u}, T, \bm{q})^T$) must be determined such that
\begin{equation}
    \bm{R}(\bm{A}^*)=\frac{V_{pvs}}{N_r}\sum_{\alpha=1}^{N_r}\begin{pmatrix}{g}^{S}_{\alpha}(\bm{A}^*)
                                     \\\bm\xi_{\alpha}{g}^{S}_{\alpha}(\bm{A}^*)
                                     \\\left[\xi_{\alpha}^2{g}^{S}_{\alpha}(\bm{A}^*)+{h}^{S}_{\alpha}(\bm{A}^*)\right]
                                     \end{pmatrix}
                          -\begin{pmatrix}\rho
                          \\\rho\bm{u}
                          \\2\rho{E}
                          \end{pmatrix}\equiv\bm{0}. \\
\label{eq:numerical_error_null}
\end{equation}
Besides the conservation constraints, additional constraint on the heat flux $\bm{q}$ can also be employed in the correction,
\begin{equation}
    \frac{V_{pvs}}{N_r}\sum_{\alpha=1}^{N_r}\left\{\bm{c}_{\alpha}^*\left[({c}_{\alpha}^*)^{2}{g}^{S}_{\alpha}(\bm{A}^*)+{h}^{S}_{\alpha}(\bm{A}^*)\right]-
    \bm{c}_{\alpha}({c}_{\alpha}^2{g}_{\alpha}+{h}_{\alpha})\right\}=-2\mathrm{Pr}\bm{q}, \\
\label{eq:heat_flux_corrected}
\end{equation}
which ensures accurate heat flux evaluation as reported in Ref.~\cite{titarev2007conservative}.

Combining the conservation constraints (Eq.~\eqref{eq:numerical_error_null}) with the heat-flux constraint (Eq.~\eqref{eq:heat_flux_corrected}) yields the nonlinear system for $\bm{A}^*$,
\begin{equation}
\bm{R}^{\prime}(\bm{A}^*)=
\frac{V_{pvs}}{N_r}\sum_{\alpha=1}^{N_r}(\begin{pmatrix}1
                           \\\bm\xi_{\alpha}
                           \\\xi_{\alpha}^2
                           \\\bm{c}_{\alpha}^*(c_{\alpha}^*)^2
                           \end{pmatrix}{g}^{S}_{\alpha}(\bm{A}^*)
                           +\begin{pmatrix}0
                           \\\bm{0}
                           \\1
                           \\\bm{c}_{\alpha}^*
                           \end{pmatrix}{h}^{S}_{\alpha}(\bm{A}^*))
                           -\begin{pmatrix}
 \rho\\
 \rho\bm{u}\\
 2\rho{E}\\
2(1-\mathrm{Pr})\bm{q}
\end{pmatrix}\equiv\bm{0}, \\
\label{eq:conservation_heatflux_corrected}
\end{equation}
which can be solved via the Newton method. 

\subsection{Algorithm}\label{sec24}
\begin{figure}[!ht]
\centering
  \subfloat{\includegraphics[width=0.9\textwidth]{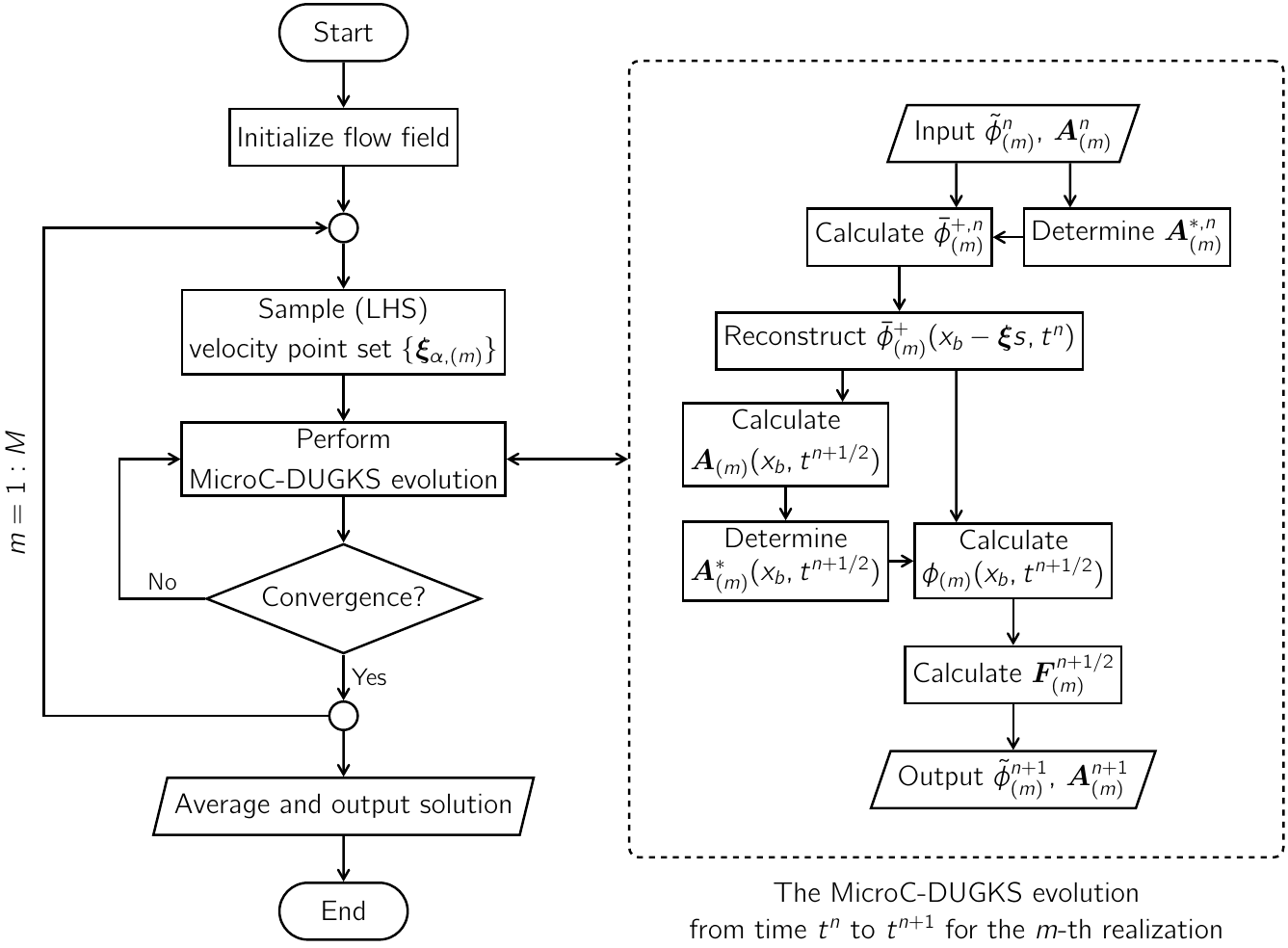}}~~
 \caption{Flow chart of the SDV-DUGKS.}
\label{Flow_chart_241}
\end{figure}

The proposed method is hereafter referred to as SDV-DUGKS for convenience. 
Its procedure is illustrated in Fig.~\ref{Flow_chart_241} and summarized as follows:
\begin{enumerate}
\item Pre-processing step
    \begin{enumerate}
        \item Set the number of independent realizations $M$.
        \item Initialize the macroscopic flow field.
        \item For each realization, sample the stochastic discrete velocity set $\{\bm{\xi}_\alpha\}$ and initialize its corresponding distribution functions $\tilde\phi$ at $t_0$.
    \end{enumerate}
\item Evolution procedure of the MicroC-DUGKS from $t_n$ to $t_{n+1}$ in per realization
    \begin{enumerate}
        \item Determine the corrected state $\bm{A}^*$ at each cell center and time $t_{n}$ by Eq.~\eqref{eq:conservation_heatflux_corrected}.
        \item Compute the auxiliary distribution $\bar\phi^+$ at each cell center and $t_n$ from $\tilde\phi$ and $\phi^{S}(\bm{A}^*)$ according to Eqs.~\eqref{eq:auxiliaryphitilde} and~\eqref{eq:auxiliaryphibar}.
        \item Reconstruct $\bar\phi^+$ at $(x_b-\bm\xi s)$ according to Eq.~\eqref{eq:phiBarinterfaceGot}.
        \item Obtain the distribution function $\bar\phi$ at $x_b$ and $t_{n+1/2}$ according to Eq.~\eqref{eq:barphievelotion}.
        \item Evaluate the conservative macroscopic variables $\bm{W}(x_b,t_{n+1/2})$ and heat flux $\bm{q}(x_b,t_{n+1/2})$ from $\bar\phi$ according to Eqs.~\eqref{eq:gBarCMoments}, ~\eqref{eq:gBarq} and~\eqref{eq:sum_macro_moments_stoca}.
        \item Determine the corrected state $\bm{A}^*$ at each cell interface and time $t_{n+1/2}$ according to Eq.~\eqref{eq:conservation_heatflux_corrected}.
        \item Calculate the original distribution function $\phi$ at $x_b$ and $t_{n+1/2}$ from $\bar\phi$ and $\phi^{S}(\bm{A}^*)$ according to Eq.~\eqref{eq:phiBarShakhov}.
        \item Calculate the microflux $\bm{F}^{n+1/2}$ across the cell interface from $\phi(x_b,\bm\xi,t_{n+1/2})$ according to Eq.~\eqref{eq:microfluxF}.
        \item Update $\tilde\phi$ at each cell center and $t_{n+1}$ from $\bm{F}^{n+1/2}$ according to Eqs.~\eqref{eq:auxiliaryphitilde} and~\eqref{eq:tildephievelution}.
    \end{enumerate}
\item Post-processing step
    \begin{enumerate}
        \item After each realization converges, evaluate the macroscopic variables at each cell center according to Eqs.~\eqref{eq:gTlideCMoments}, ~\eqref{eq:gTlideq} and~\eqref{eq:sum_macro_moments_stoca}.
        \item Compute the average of these results over all $M$ realizations and output the final macroscopic flow field according to Eq.~\eqref{eq:average_M}.
    \end{enumerate}
\end{enumerate}

\section{Numerical results and discussions}\label{sec3}
This section presents a performance validation of SDV-DUGKS, with comparisons against the original DUGKS, using several test problems: $1$D shock structure, $2$D lid-driven cavity flow, and supersonic flow around a square cylinder. The gas constant and Prandtl number are set as $R=0.5$ and $\mathrm{Pr}=2/3$, respectively.
The time step $ \Delta t $ is determined by the Courant-Friedrichs-Lewy (CFL) condition~\cite{de2013courant},
\begin{equation}
   \Delta t=\beta\frac{\Delta{x}}{|\bm\xi|_{\operatorname*{max}}+|\bm{u}|_{\operatorname*{max}}},  \\
\label{eq:CFLnumber}
\end{equation}
where $\beta \in (0, 1]$ is the CFL number, set to $\beta = 0.5$ for all cases.
In addition, the relative global error is defined as:
\begin{equation}
   E=\frac{\sqrt{\sum_{j=1}^{N_c}\|I_j(\bm{a})-I_j^{*}(\bm{a})\|^2}}{\sqrt{\sum_{j=1}^{N_c}\|I_j^{*}(\bm{a})\|^2}},  \\
\label{eq:RelativeL2error}
\end{equation}
where $I^*$ denotes the reference solution, $I(\cdot)$ the numerical result, and $N_c$ the total number of cells. 
Note that the CPU hardware is an Intel Xeon Gold 6348 CPU @ 2.60 GHz processor.

In the simulations, the distribution functions are initialized as the Maxwellian distribution corresponding to the given macroscopic conditions.
The solution is then advanced in time until a steady-state is reached, with convergence judged for the original DUGKS by the average relative change in each conserved macroscopic variable between successive steps of less than $1.0 \times 10^{-8}$, i.e.,
\begin{equation}
\varepsilon_{k} = \frac{ \sum_{j=1}^{N_c} | W_{j,k}^{n+1} - W_{j,k}^n | }{ \sum_{j=1}^{N_c} | W_{j,k}^n | } < 1.0 \times 10^{-8}, \quad\quad \forall \, k \in \{1, 2, \dots, D+2\}, \\
\label{eq:convergence_criterion}
\end{equation}
where the subscript $k$ denotes the $k$-th component of the vector $\bm{W}_j=(\rho_j, \rho_j \bm{u}_j, \rho_j{E}_j)^T $, and $N_c$ the total number of cells.

\subsection{$1\mathrm{D}$ shock structure}\label{sec31}
In this subsection, the $1\mathrm{D}$ argon shock structure from low to high Mach numbers is considered to provide a feasibility validation of the proposed SDV-DUGKS.
The specific heat ratio for argon is $\gamma = 5/3$.
The initial condition comprises two uniform regions, upstream $(x \le 0)$ and downstream $(x > 0)$, separated by a discontinuity at $x=0$, with the respective states given by $(\rho_1, u_1, T_1)$ and $(\rho_2, u_2, T_2)$.
The dynamic viscosity follows $\mu = \mu_{\mathrm{ref}} (T / T_1)^\omega$, where $\mu_{\mathrm{ref}}$ represents the reference viscosity at the reference temperature $T_{\mathrm{ref}} = T_1$, and the exponent $\omega$ is a constant depending on the inter-molecular interaction model~\cite{bird1994molecular}.
The reference mean free path $\lambda_\mathrm{ref}$ is related to the reference viscosity $\mu_\mathrm{ref}$ as~\cite{bird1994molecular}
\begin{equation}
   \lambda_\mathrm{ref}=\frac{2(7-2\omega)(5-2\omega)\mu_\mathrm{ref}}{15\rho_\mathrm{ref}(2\pi R T_\mathrm{ref})^{1/2}},              \\
   \label{eq:ref_mfp}
\end{equation}
where $\rho_\mathrm{ref}=\rho_1$ is the reference density.
In the simulations, the upstream quantities are set as $\rho_1 = 1$, $T_1 = 1$, $\lambda_1 = 1$, with the velocity $u_1 = \mathrm{Ma}\sqrt{\gamma R T_1}$ for a given upstream Mach number $\mathrm{Ma}$.
The downstream state $(\rho_2, u_2, T_2)$ follows from the Rankine–Hugoniot conditions~\cite{harris2004introduction}:
\begin{subequations}
    \begin{equation}
    \frac{u_2}{\sqrt{\gamma R T_2}}=\sqrt{\frac{\mathrm{Ma}^2(\gamma-1)+2}{2\gamma\mathrm{Ma}^2-(\gamma-1)}}, \\
    \label{eq:R_H_U}
    \end{equation}
    \begin{equation}
    \frac{\rho_2}{\rho_1}=\frac{(\gamma+1)\mathrm{Ma}^2}{(\gamma-1)\mathrm{Ma}^2+2},\\
    \label{eq:R_H_rho}
    \end{equation}
    \begin{equation}
    \frac{T_2}{T_1}=\frac{(1+\frac{\gamma-1}{2}\mathrm{Ma}^2)(\frac{2\gamma}{\gamma-1}\mathrm{Ma}^2-1)}{\mathrm{Ma}^2(\frac{2\gamma}{\gamma-1}+\frac{\gamma-1}{2})}.\\
    \label{eq:R_H_T}
    \end{equation}
    \label{eq:R_H_conditions}%
\end{subequations}
The computational domain $[-25\lambda_1, \, 25\lambda_1]$ is discretized with $100$ uniform cells, a resolution sufficient to capture the shock structure and produce grid-independent results.
For all test cases in this subsection, the reference solution is computed using the original DUGKS with a uniform discretization of $401$ velocity points in the velocity space domain $[-15, 15]$ based on the Newton–Cotes quadrature rule~\cite{hayes1970proof}.

\begin{figure}[!ht]
\centering
  \subfloat{}{\includegraphics[width=0.4\textwidth]{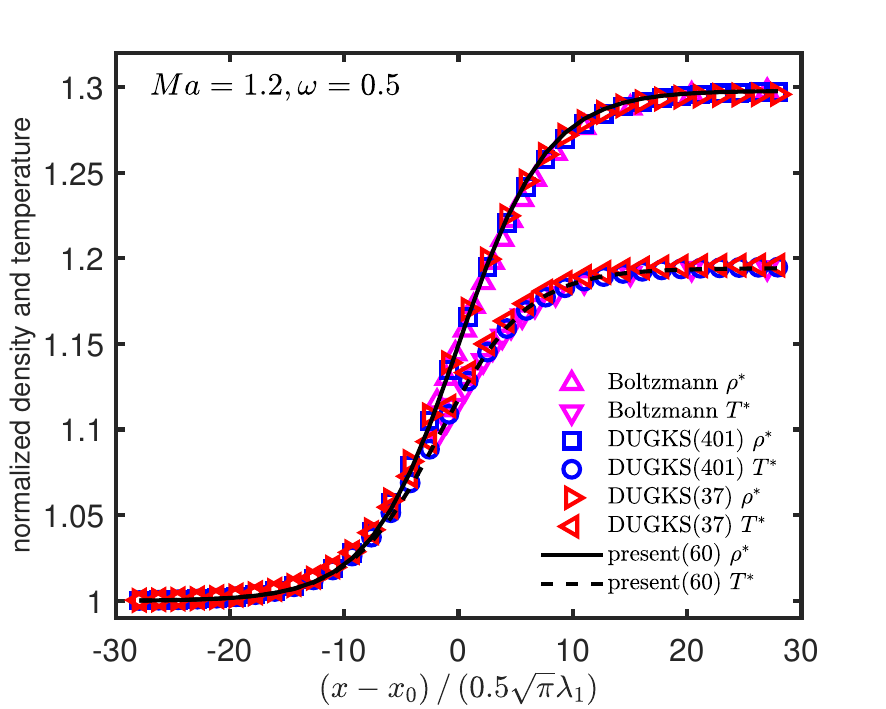}}~~
      \hspace{6mm}
  \subfloat{}{\includegraphics[width=0.4\textwidth]{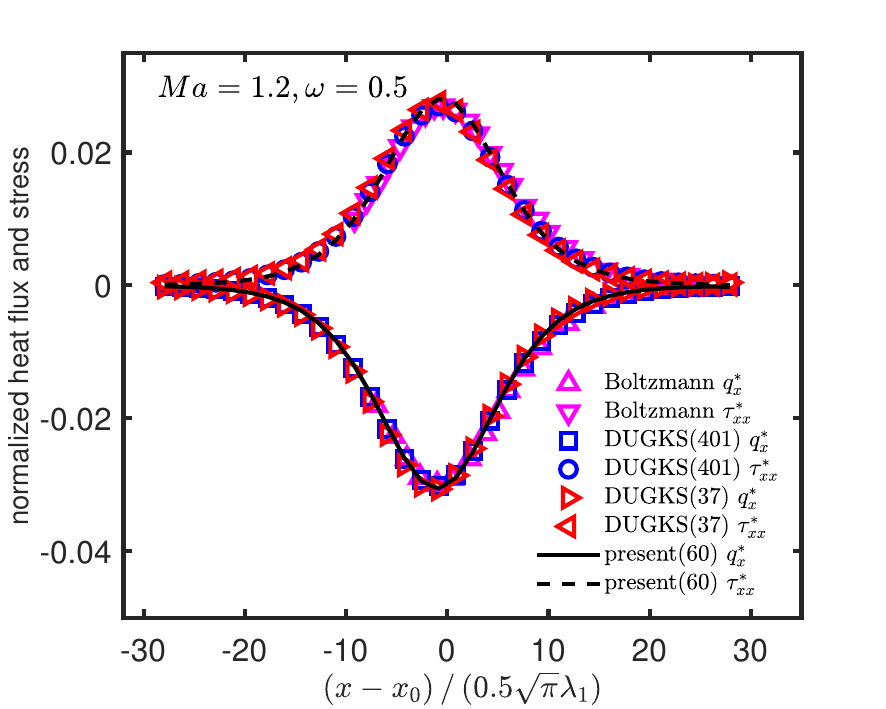}}~~
 \caption{Shock structure with $\mathrm{Ma} = 1.2$ and $\omega = 0.5$ ($\mu_{\text{ref}} = 0.5539$): (left) density ($\rho^*$) and temperature ($T^*$); (right) stress ($\tau_{xx}^*$) and heat flux ($q_x^*$).}
\label{shock_structure311}
\end{figure}

\begin{figure}[!ht]
\centering
  \subfloat{}{\includegraphics[width=0.4\textwidth]{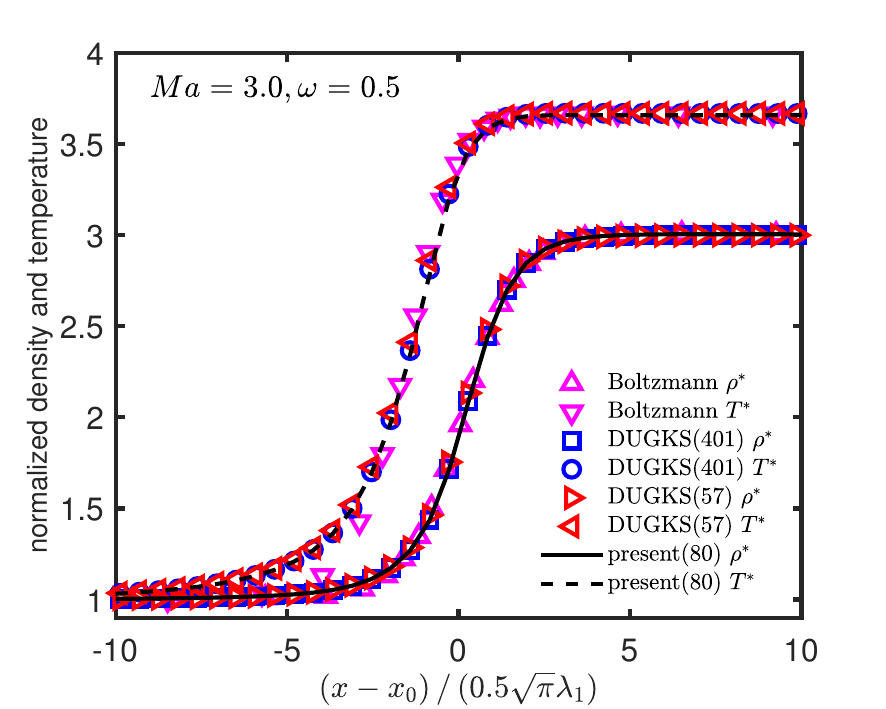}}~~
      \hspace{6mm}
  \subfloat{}{\includegraphics[width=0.4\textwidth]{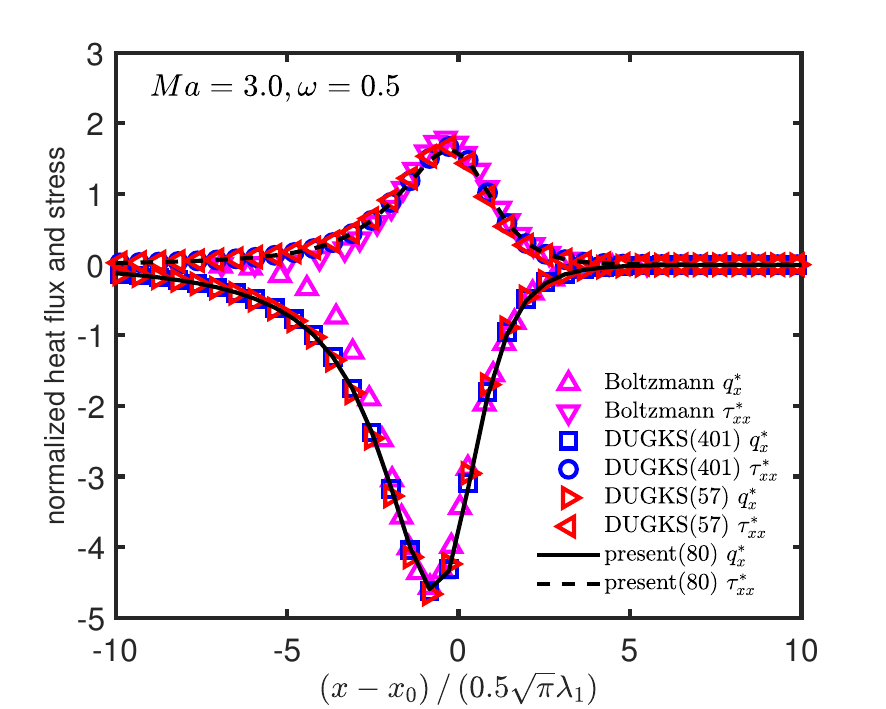}}~~
 \caption{Shock structure with $\mathrm{Ma} = 3.0$ and $\omega = 0.5$ ($\mu_{\text{ref}} = 0.5539$): (left) density ($\rho^*$) and temperature ($T^*$); (right) stress ($\tau_{xx}^*$) and heat flux ($q_x^*$).}
\label{shock_structure312}
\end{figure}

\begin{figure}[!ht]
\centering
  \subfloat{}{\includegraphics[width=0.4\textwidth]{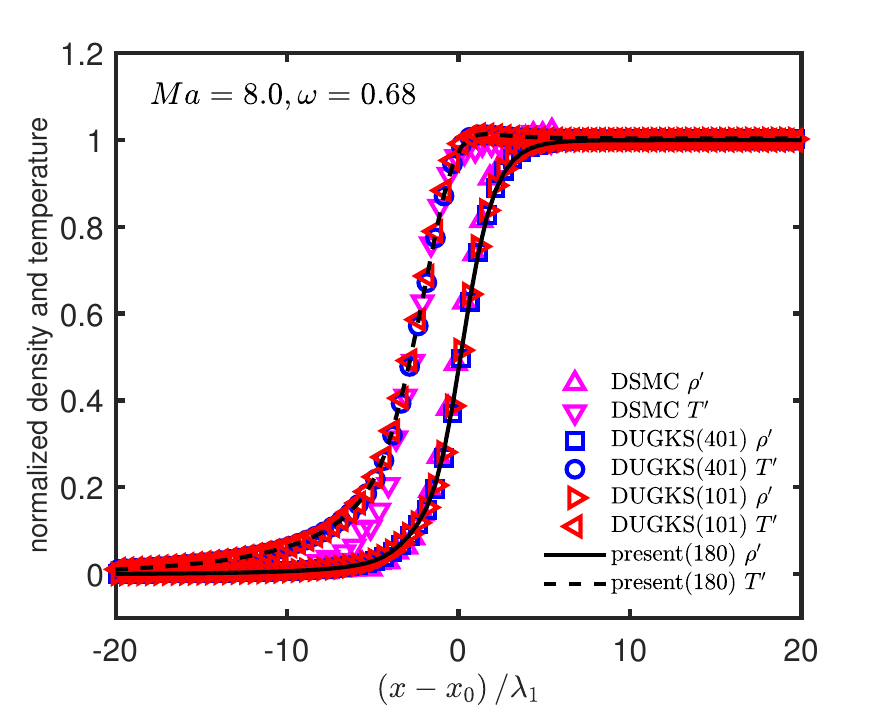}}~~
      \hspace{6mm}
  \subfloat{}{\includegraphics[width=0.4\textwidth]{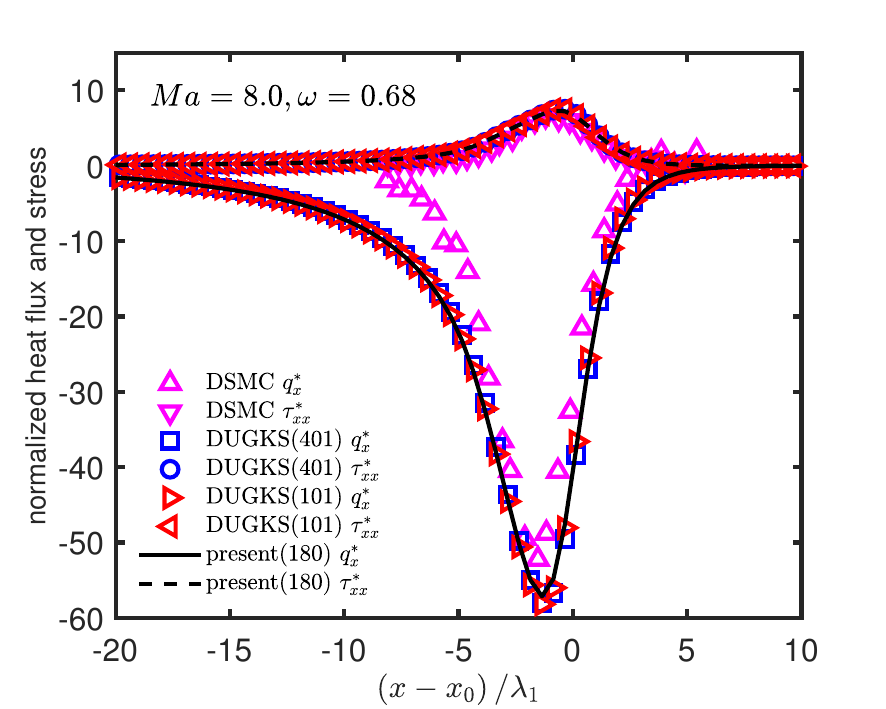}}~~
 \caption{Shock structure with $\mathrm{Ma} = 8.0$ and $\omega = 0.68$ ($\mu_{\text{ref}} = 0.5539$): (left) density ($\rho^{\prime}$) and temperature ($T^{\prime}$); (right) stress ($\tau_{xx}^*$) and heat flux ($q_x^*$).}
\label{shock_structure313}
\end{figure}

\begin{table}[!ht]
\caption{Computational cost and relative global errors of DUGKS and SDV-DUGKS for $1\mathrm{D}$ shock structure at various Mach numbers.}
\centering
\setlength{\tabcolsep}{2.8mm}
\renewcommand\arraystretch{1.4}
\begin{tabular}{c c c c c c c c c}
 \toprule
 \multirow{2}{*}{Case} & \multirow{2}{*}{Method} & \multirow{2}{*}{\makecell[c]{Velocity \\ Point}} & \multirow{2}{*}{$M$}
 & \multirow{2}{*}{\makecell[c]{Memory \\ (MB)}}  & \multicolumn{4}{c}{$E(\%)$} \\
 \cmidrule(lr){6-9}
 \multirow{2}{*}{} & \multirow{2}{*}{} & \multirow{2}{*}{} & \multirow{2}{*}{} & \multirow{2}{*}{} & $\rho$  & $T$ & $q_x$  & $\tau_{xx}$  \\
 \midrule
 \multirow{2}{*}{\makecell[c]{$\mathrm{Ma} = 1.2$}} & {DUGKS} & 37 & 1 & 4.152  & 0.176 & 0.236 & 3.979 & 3.512 \\
 \cmidrule(lr){2-9}
 \multirow{2}{*}{} & {SDV-DUGKS} & 60 & 20 & 4.461  & 0.184 & 0.115 & 2.905 & 3.856  \\
 \midrule
 \multirow{2}{*}{\makecell[c]{$\mathrm{Ma} = 3.0$}} & {DUGKS} & 57 & 1 & 4.410  & 0.346 & 0.350 & 2.844 & 3.203 \\
 \cmidrule(lr){2-9}
 \multirow{2}{*}{} & {SDV-DUGKS} & 80 & 20 & 4.746  & 0.138 & 0.274 & 1.325 & 1.213  \\
 \midrule
 \multirow{2}{*}{\makecell[c]{$\mathrm{Ma} = 8.0$}} & {DUGKS} & 101 & 1 & 4.906  & 0.603 & 0.532 & 2.469 & 2.571 \\
 \cmidrule(lr){2-9}
 \multirow{2}{*}{} & {SDV-DUGKS} & 180 & 20 & 5.566  & 0.530 & 0.528 & 2.462 & 2.672  \\
 \bottomrule
\end{tabular}
\label{tab311}
\end{table}

\begin{table}[!ht]
\caption{Relative global errors of macroscopic quantities for the $1\mathrm{D}$ shock structure at $\mathrm{Ma} = 3.0$, computed by SDV-DUGKS with different numbers of realizations $M$.}
\centering
\setlength{\tabcolsep}{3.5mm}
\renewcommand\arraystretch{1.2}
\begin{tabular}{c c c c c c}
 \toprule
 \multirow{2}{*}{\makecell[c]{Velocity \\ Point}} & \multirow{2}{*}{$M$}  & \multicolumn{4}{c}{$E(\%)$} \\
 \cmidrule(lr){3-6}
  & & $\rho$ & $T$ & $q_x$ & $\tau_{xx}$ \\
 \midrule
 \multirowcell{5}{80} & $1$ & 0.693 & 1.283 & 9.869 & 4.842 \\
 \cmidrule(lr){2-6}
  & $5$ & 0.344 & 0.483 & 5.751 & 2.797 \\
 \cmidrule(lr){2-6}
  & $10$ & 0.248 & 0.426 & 2.334 & 2.087 \\
 \cmidrule(lr){2-6}
  & $20$ & 0.138 & 0.274 & 1.325 & 1.213 \\
 \bottomrule
\end{tabular}
\label{tab312}
\end{table}

First, we consider the hard-sphere model ($\omega = 0.5$), which has been studied previously by numerically solving the full Boltzmann equation~\cite{ohwada1993structure} and by applying the UGKS~\cite{xu2010unified} and DUGKS~\cite{guo2015discrete}.
For the cases at $\mathrm{Ma} = 1.2$ and $3.0$, the original DUGKS uniformly discretizes the velocity space domains $[-5, 5]$ and $[-8, 8]$ into $37$ and $57$ discrete velocity points, respectively, based on the Newton–Cotes quadrature rule. 
Meanwhile, SDV-DUGKS employs LHS to sample $60$ and $80$ velocity points within the corresponding domains.
Each realization is advanced until the average relative change in each conserved macroscopic variable between successive steps below $10^{-6}$, i.e.,
\begin{equation}
\varepsilon^{\prime}_{k} = \frac{ \sum_{j=1}^{N_c} | W_{j,k}^{n+1} - W_{j,k}^n | }{ \sum_{j=1}^{N_c} | W_{j,k}^n | } < 1.0 \times 10^{-6}, \quad\quad \forall \, k \in \{1, 2, \dots, D+2\}.
\label{eq:convergence_criterion_MC}
\end{equation}
The results presented are averaged over $M = 20$ such independent realizations.
Figures~\ref{shock_structure311} and~\ref{shock_structure312} present the profiles of normalized macroscopic quantities at $\mathrm{Ma} = 1.2$ and $3.0$, respectively, for both DUGKS and SDV-DUGKS. 
The normalized quantities are defined as follows: density $\rho^* = \rho/\rho_\mathrm{ref}$, temperature $T^* = T/T_\mathrm{ref}$, heat flux $q_x^* = q_x/[p_\mathrm{ref} (2RT_\mathrm{ref})^{1/2}]$, and shear stress $\tau_{xx}^* = \tau_{xx}/p_\mathrm{ref}$, where $p_\mathrm{ref} = \rho_\mathrm{ref} R T_\mathrm{ref}$. 
Here the shock location $x_0$ is chosen such that $\rho(x_0) = (\rho_1 + \rho_2)/2$.
The results are compared with the Boltzmann and reference DUGKS solutions.
At $\mathrm{Ma} = 1.2$, the density, temperature, shear stress, and heat flux predicted by SDV-DUGKS show excellent agreement with the DUGKS results, as well as with the reference DUGKS solution and the Boltzmann solution.
At $\mathrm{Ma} = 3.0$, the SDV-DUGKS results are in good agreement with those of DUGKS and the reference DUGKS solution.
While the density and stress profiles from both methods agree well with the Boltzmann solution, their predictions for temperature and heat flux show clear deviations in the upstream region.
In particular, compared to the Boltzmann solution, the temperature profiles predicted by both DUGKS and SDV-DUGKS rise earlier in the upstream region, with a corresponding earlier decrease observed in the heat flux profiles.
These deviations are attributed to the use of a single relaxation time in the Shakhov collision model~\cite{guo2015discrete,xu2021modeling,yuan2022capturing}, which is beyond the scope of the present numerical investigation.

We next test the shock structure at $\mathrm{Ma} = 8$ with $\omega = 0.68$, which has been studied numerically based on the DSMC~\cite{bird1970aspects} and the original DUGKS~\cite{guo2015discrete}.
For this case, the original DUGKS uniformly discretizes the velocity space domain $[-15, 15]$ into $101$ discrete velocity points based on the Newton–Cotes quadrature rule, whereas SDV-DUGKS employs LHS with $180$ points within the same domains.
The SDV-DUGKS simulation here adopts the same convergence criterion as specified in Eq.~\ref{eq:convergence_criterion_MC}, and the same ensemble size of $M = 20$.
Figure~\ref{shock_structure313} presents the profiles of the normalized density $\rho^{\prime}=(\rho-\rho_1)/(\rho_2-\rho_1)$, temperature $T^{\prime}=(T-T_1)/(T_2-T_1)$, heat flux $q_x^*$, and shear stress $\tau_{xx}^*$ for both DUGKS and SDV-DUGKS, compared with the DSMC and DUGKS results~\cite{bird1970aspects,guo2015discrete}.
The predictions of SDV-DUGKS align closely with those of the original DUGKS and the reference DUGKS solution. 
Compared to the DSMC results, the density and stress profiles from both SDV-DUGKS and DUGKS agree well, while more pronounced deviations are observed in their predictions for temperature and heat flux within the upstream region. 
This behavior, consistent with the $\mathrm{Ma}=3.0$ case, is also attributed to the single-relaxation-time approximation in the Shakhov collision model.
Moreover, the peak values of the heat flux profile are quite close across the three methods.

In addition, a detailed comparison of the computational cost and accuracy between DUGKS and SDV-DUGKS across all test cases is summarized in Table~\ref{tab311}. 
For the one-dimensional cases, SDV-DUGKS achieves accuracy comparable to the original DUGKS but at a higher computational cost. 
This is expected, as Monte Carlo integration is acknowledged to be less effective in low-dimensional settings. 
Therefore, these one-dimensional simulations serve primarily as a preliminary validation of the feasibility of SDV-DUGKS. 
Furthermore, Table~\ref{tab312} illustrates the significance of averaging over multiple realizations. 
With limited samples and lower integration accuracy, the result from any single realization appears unsatisfactory. 
However, the final SDV-DUGKS result, obtained by averaging over all $M$ realizations, shows progressively improving accuracy as $M$ increases, eventually agreeing well with the reference solution.

\subsection{$2\mathrm{D}$ lid-driven cavity flow}\label{sec32}
\begin{figure}[!ht]
\centering
  \subfloat{\includegraphics[width=0.25\textwidth]{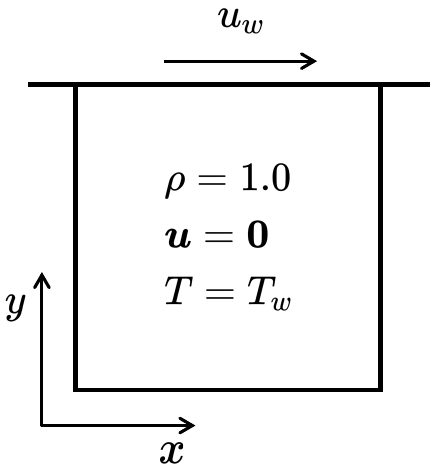}}~~
 \caption{The schematic of the $2\mathrm{D}$ lid-driven cavity flow.}
\label{2D_cavity}
\end{figure}

The two-dimensional lid-driven cavity flow is a classic benchmark for validating numerical methods in fluid mechanics~\cite{kuhlmann2018lid}.
To demonstrate the capability of the SDV-DUGKS for simulating subsonic rarefied flow, we consider this classic problem at Knudsen numbers $\mathrm{Kn} = 1.0$ and $10.0$.
As illustrated in Fig.~\ref{2D_cavity}, the cavity is a square with side length $L = 1$. 
All walls are maintained at a constant temperature $T_w = 1.099$, while the lid moves in the positive $x$-direction with velocity $u_w = 0.274$.
Initially, the fluid is at rest with density $\rho_0 = \rho_\mathrm{ref} = 1.0$ and temperature $T_0 = T_w$. 
The dynamic viscosity is given by $\mu = \mu_{\mathrm{ref}} (T / T_{\mathrm{ref}})^\omega$ with $\omega = 0.81$ and $T_\mathrm{ref} =1.0$, where the reference viscosity $\mu_{\mathrm{ref}}$ is expressed as~\cite{bird1994molecular}
\begin{equation}
    \mu_\text{ref}={\frac{15\rho_\text{ref}(2\pi R T_\text{ref})^{1/2}}{2(7-2\omega)(5-2\omega)}}\mathrm{Kn}{L}. \\
\label{eq:referenceviscosity}
\end{equation}
Diffuse reflection boundary conditions~\cite{guo2013discrete,li2005application} are applied to all boundaries.
A uniform Cartesian grid of $60 \times 60$ cells is used to discretize the computational domain $[0,1]^2$ for both the original DUGKS and the present SDV-DUGKS.

\begin{figure}[!ht]
\centering
  \subfloat[$Kn = 1.0$]{\includegraphics[width=0.4\textwidth]{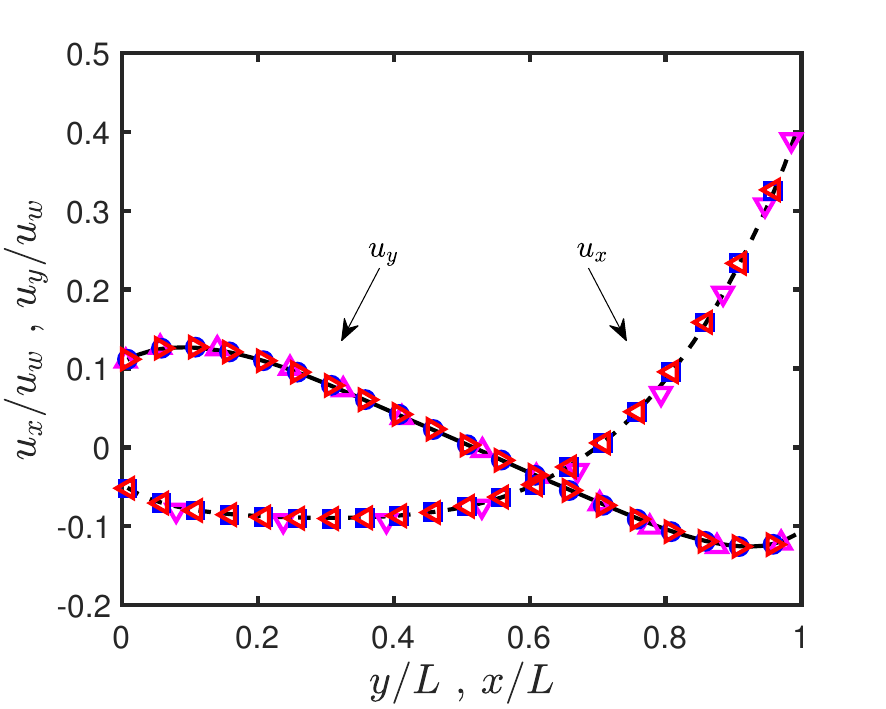}}~~
      \hspace{6mm}
  \subfloat[$Kn = 10.0$]{\includegraphics[width=0.4\textwidth]{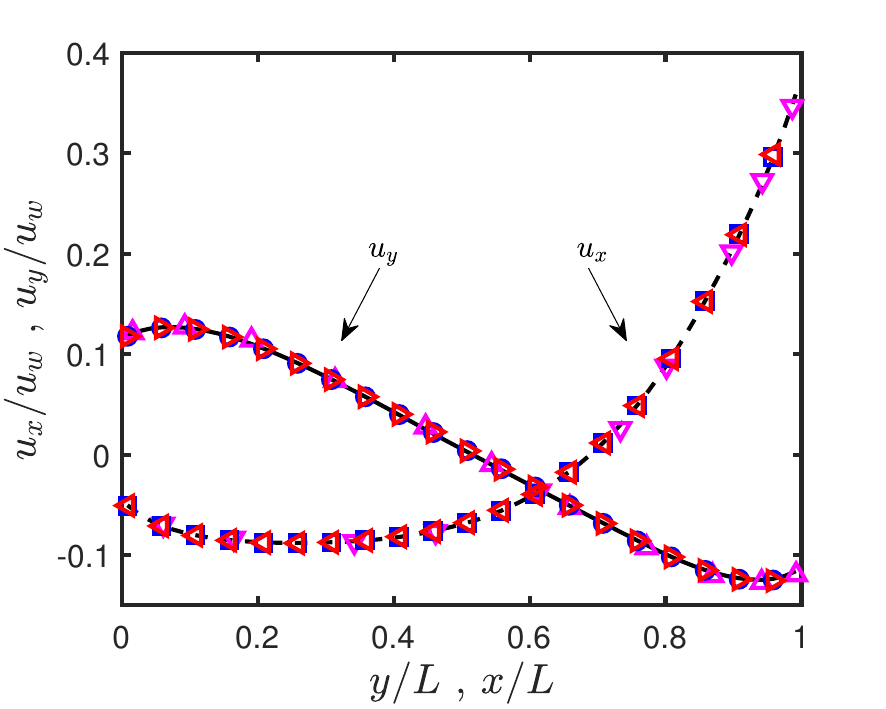}}~~
 \caption{$u_x$-velocity profile along the mid-vertical line and $u_y$-velocity profile along the mid-horizontal line at $Kn = 1.0$ $(a)$ and $Kn = 10.0$ $(b)$. Magenta open triangle: DSMC results~\cite{venugopal2015unified}; open circle and square: Reference solution; red open triangle: DUGKS results; black solid and dashed line: SDV-DUGKS results.}
\label{MicrocavityFlow_u_322}
\end{figure}

\begin{figure}[!ht]
\centering
  \subfloat[$Kn = 1.0$]{\includegraphics[width=0.4\textwidth]{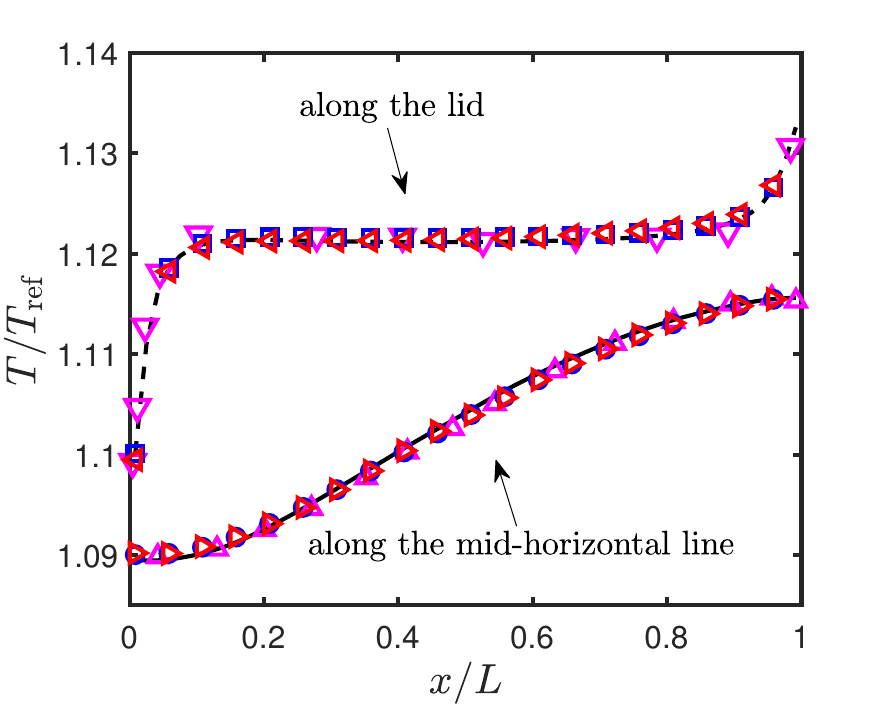}}~~
      \hspace{6mm}
  \subfloat[$Kn = 10.0$]{\includegraphics[width=0.4\textwidth]{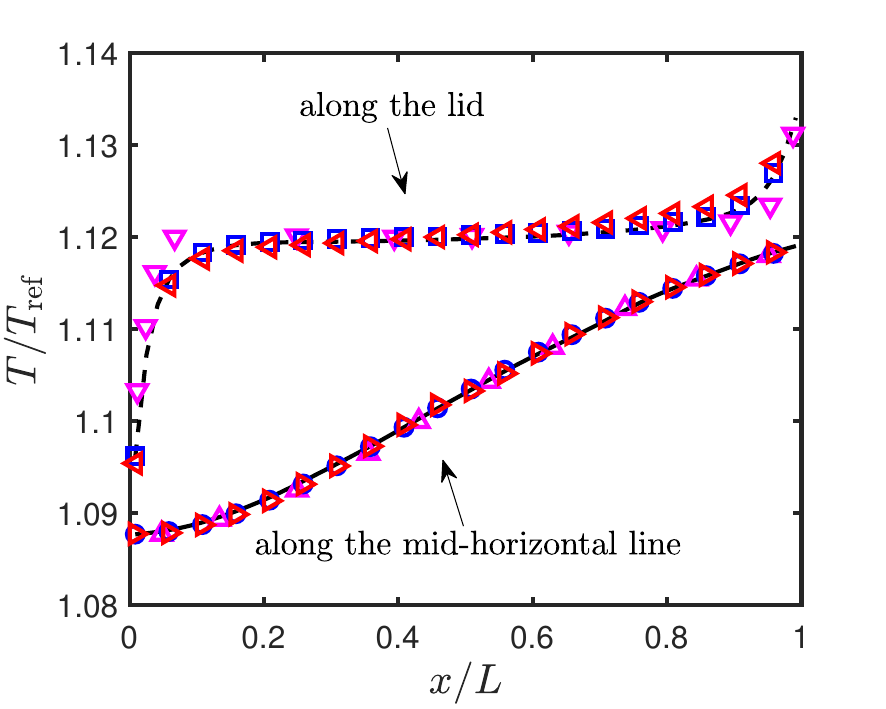}}~~
 \caption{Temperature profile along the mid-horizontal line and the lid at $Kn = 1.0$ $(a)$ and $Kn =10.0$ $(b)$. Magenta open triangle: DSMC results~\cite{venugopal2015unified}; open circle and square: Reference solution; red open triangle: DUGKS results; black solid and dashed line: SDV-DUGKS results.}
\label{MicrocavityFlow_T_323}
\end{figure}

\begin{table}[!ht]
\caption{Computational cost and relative global errors of DUGKS and SDV-DUGKS for $2\mathrm{D}$ lid-driven cavity flow at various Knudsen numbers.}
\centering
\setlength{\tabcolsep}{2.8mm}
\renewcommand\arraystretch{1.4}
\begin{tabular}{c c c c c c c c c}
 \toprule
 \multirow{2}{*}{Case} & \multirow{2}{*}{Method} & \multirow{2}{*}{\makecell[c]{Velocity \\ Point}} & \multirow{2}{*}{$M$}
 & \multicolumn{2}{c}{Memory}  & \multicolumn{3}{c}{$E(\%)$} \\
 \cmidrule(lr){5-6} \cmidrule(lr){7-9}
  & & & & Total (GB) & Ratio & $u_x$ & $u_y$ & $T$ \\
 \midrule
 \multirow{2}{*}{$\mathrm{Kn} = 1.0$} & DUGKS & $81^2$ & 1 & 3.319 & 1.0  & 0.729 & 0.232 & 0.014 \\
 \cmidrule(lr){2-9}
 & SDV-DUGKS & 1000 & 100 & 0.497 & $\sim 1/7$  & 0.918 & 0.895 & 0.034 \\
  \midrule
 \multirow{2}{*}{$\mathrm{Kn} = 10.0$} & DUGKS & $101^2$ & 1 & 5.146 & 1.0 & 0.775 & 0.368 & 0.022 \\
 \cmidrule(lr){2-9}
 & SDV-DUGKS & 1000 & 200 & 0.496 & $\sim 1/10$ & 0.729 & 0.710 & 0.018 \\
 \bottomrule
\end{tabular}
\label{tab321}
\end{table}

\begin{figure}[!ht]
\centering
  \subfloat[]{\includegraphics[width=0.35\textwidth]{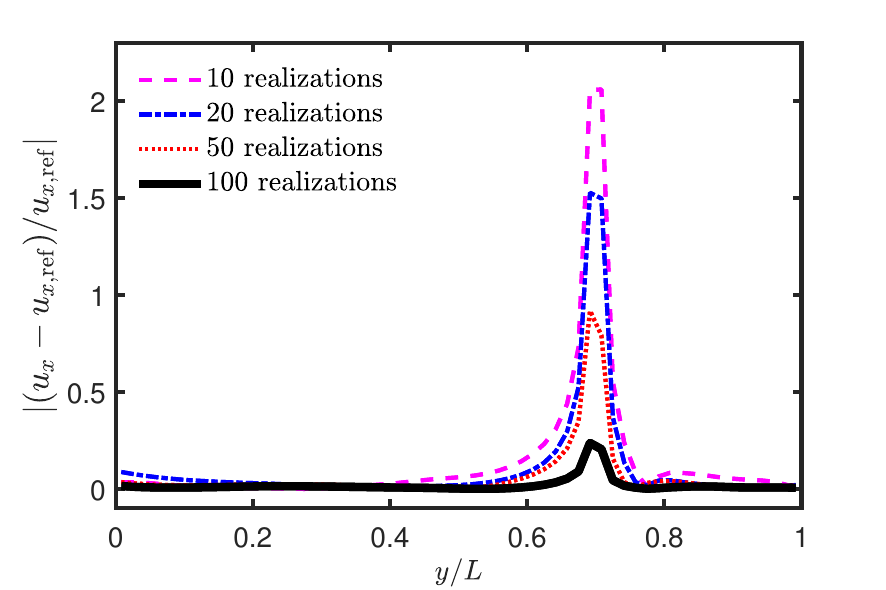}}~~
      \hspace{1mm}
  \subfloat[]{\includegraphics[width=0.35\textwidth]{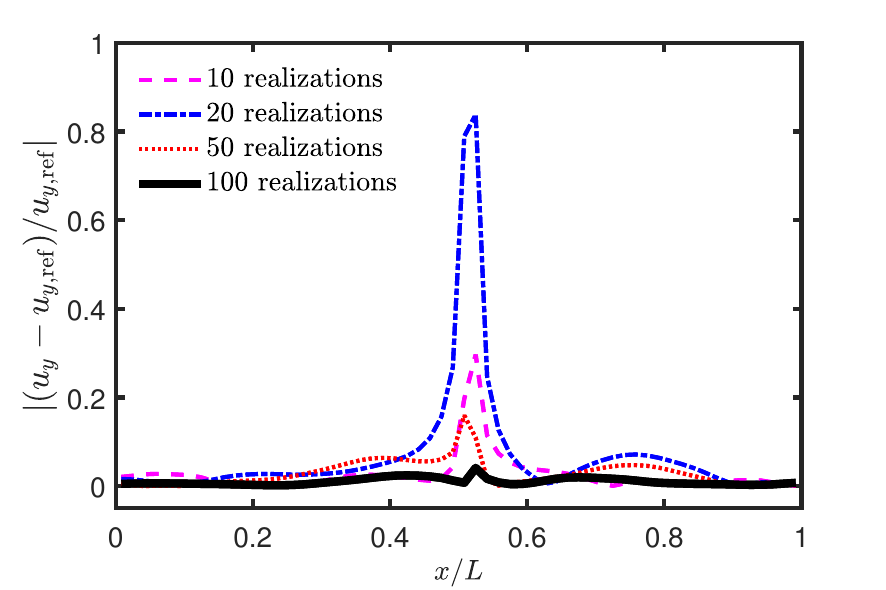}}~~
      \hspace{1mm}
  \subfloat[]{\includegraphics[width=0.35\textwidth]{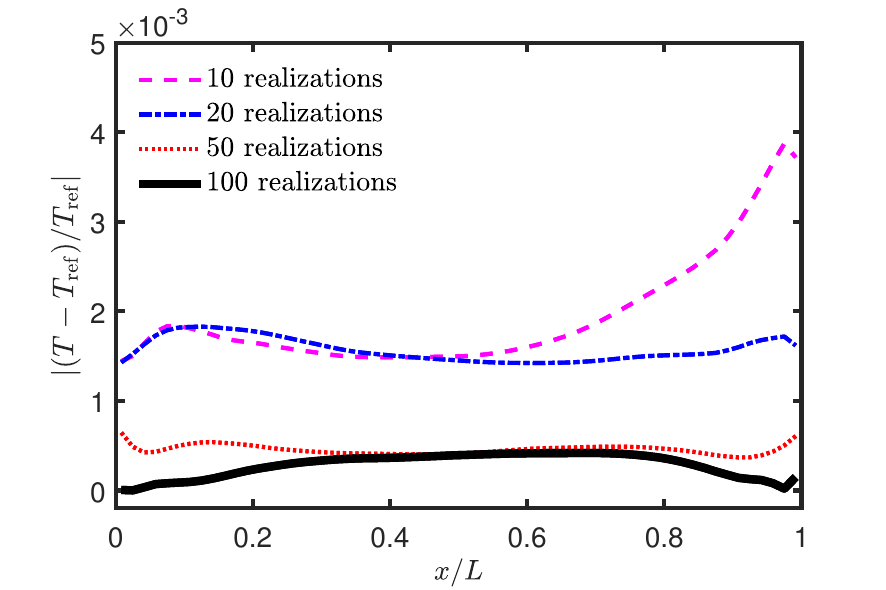}}~~
 \caption{Local relative errors in macroscopic quantities computed by SDV-DUGKS with different numbers of realizations $M$ at $\mathrm{Kn} = 1.0$. (a) $u_x$-velocity along the vertical centerline; (b) $u_y$-velocity along the horizontal centerline; (c) temperature along the lid.}
\label{comparision_314_1}
\end{figure}

\begin{figure}[!ht]
\centering
  \subfloat[]{\includegraphics[width=0.35\textwidth]{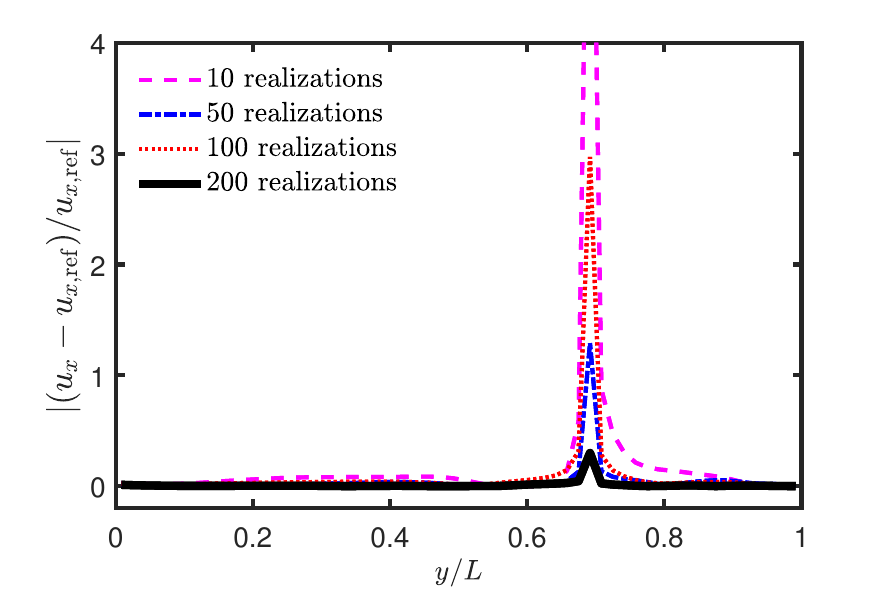}}~~
      \hspace{1mm}
  \subfloat[]{\includegraphics[width=0.35\textwidth]{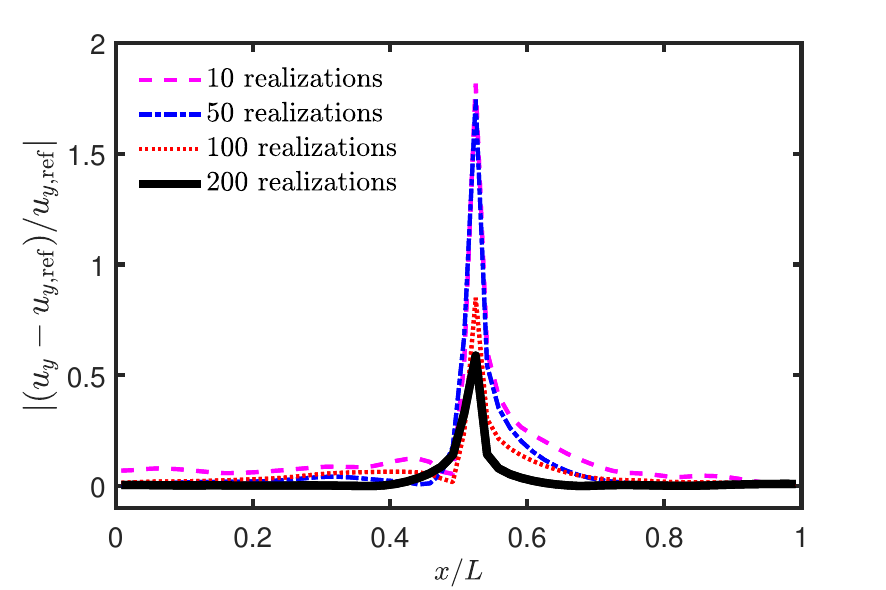}}~~
      \hspace{1mm}
  \subfloat[]{\includegraphics[width=0.35\textwidth]{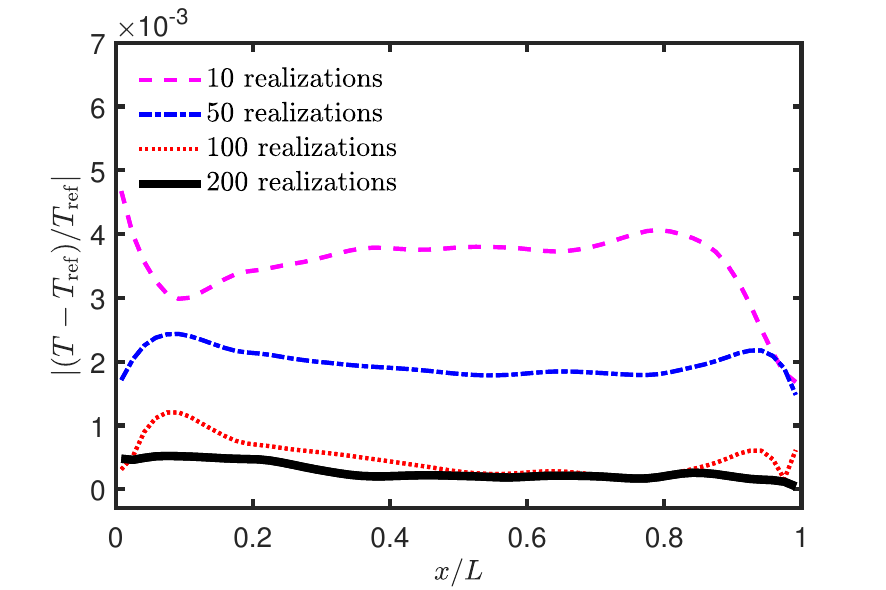}}~~
 \caption{Local relative errors in macroscopic quantities computed by SDV-DUGKS with different numbers of realizations $M$ at $\mathrm{Kn} = 10.0$. (a) $u_x$-velocity along the vertical centerline; (b) $u_y$-velocity along the horizontal centerline; (c) temperature along the lid.}
\label{comparision_314_10}
\end{figure}

\begin{table}[!ht]
\caption{Relative global errors of macroscopic quantities in the $2\mathrm{D}$ lid-driven cavity flow computed by SDV-DUGKS with different numbers of realizations $M$ at $\mathrm{Kn} = 1.0$.}
\centering
\setlength{\tabcolsep}{3.5mm}
\renewcommand\arraystretch{1.2}
\begin{tabular}{c c c c c}
 \toprule
 \multirow{2}{*}{\makecell[c]{Velocity \\ Point}} & \multirow{2}{*}{$M$}  & \multicolumn{3}{c}{$E(\%)$} \\
 \cmidrule(lr){3-5}
  & & $u_x$ & $u_y$ & $T$ \\
 \midrule
 \multirowcell{5}{1000} & $10$ & 4.405 & 3.050 & 0.298 \\
 \cmidrule(lr){2-5}
  & $20$ & 2.777 & 2.929 & 0.177 \\
 \cmidrule(lr){2-5}
  & $50$ & 1.915 & 2.009 & 0.094 \\
 \cmidrule(lr){2-5}
  & $100$ & 0.918 & 0.895 & 0.034 \\
 \bottomrule
\end{tabular}
\label{tab322}
\end{table}

\begin{table}[!ht]
\caption{Relative global errors of macroscopic quantities in the $2\mathrm{D}$ lid-driven cavity flow computed by SDV-DUGKS with different numbers of realizations $M$ at $\mathrm{Kn} = 10.0$.}
\centering
\setlength{\tabcolsep}{3.5mm}
\renewcommand\arraystretch{1.2}
\begin{tabular}{c c c c c}
 \toprule
 \multirow{2}{*}{\makecell[c]{Velocity \\ Point}} & \multirow{2}{*}{$M$}  & \multicolumn{3}{c}{$E(\%)$} \\
 \cmidrule(lr){3-5}
  & & $u_x$ & $u_y$ & $T$ \\
 \midrule
 \multirowcell{5}{1000} & $10$ & 5.829 & 5.952 & 0.116 \\
 \cmidrule(lr){2-5}
  & $50$ & 2.720 & 2.538 & 0.103 \\
 \cmidrule(lr){2-5}
  & $100$ & 2.356 & 2.555 & 0.099 \\
 \cmidrule(lr){2-5}
  & $200$ & 0.729 & 0.710 & 0.018 \\
 \bottomrule
\end{tabular}
\label{tab323}
\end{table}

\begin{figure}[!ht]
\centering
  \subfloat{\includegraphics[width=1.1\textwidth]{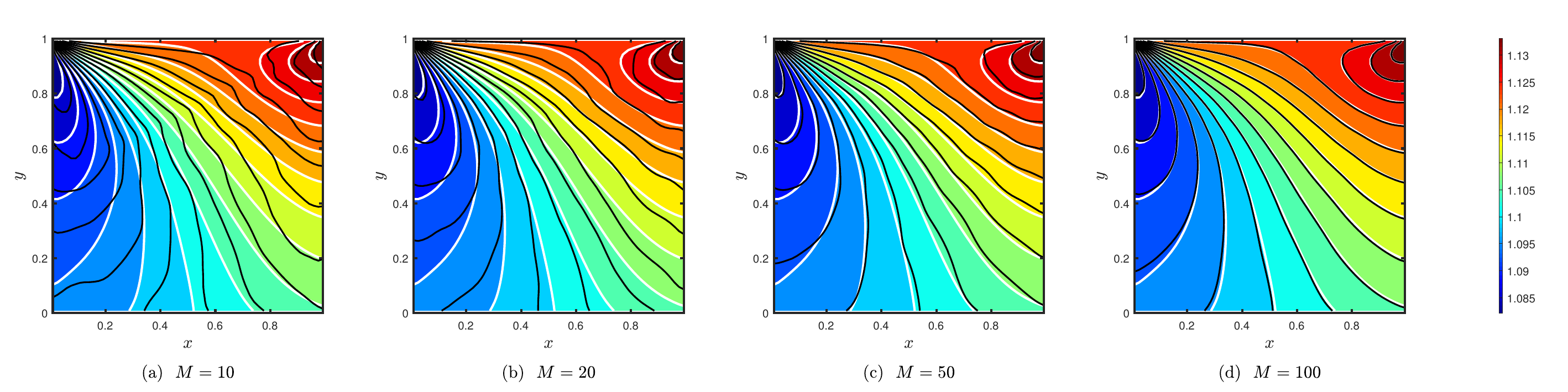}}~~
 \caption{The temperature contour for the $2\mathrm{D}$ lid-driven cavity flow at $\mathrm{Kn} = 1.0$. White solid lines with the colored background: Reference data; black solid lines: SDV-DUGKS results. (a) $M=10$; (b) $M=20$; (c) $M=50$; (d) $M=100$.}
\label{comparison_T_326}
\end{figure}

\begin{figure}[!ht]
\centering
  \subfloat{\includegraphics[width=1.1\textwidth]{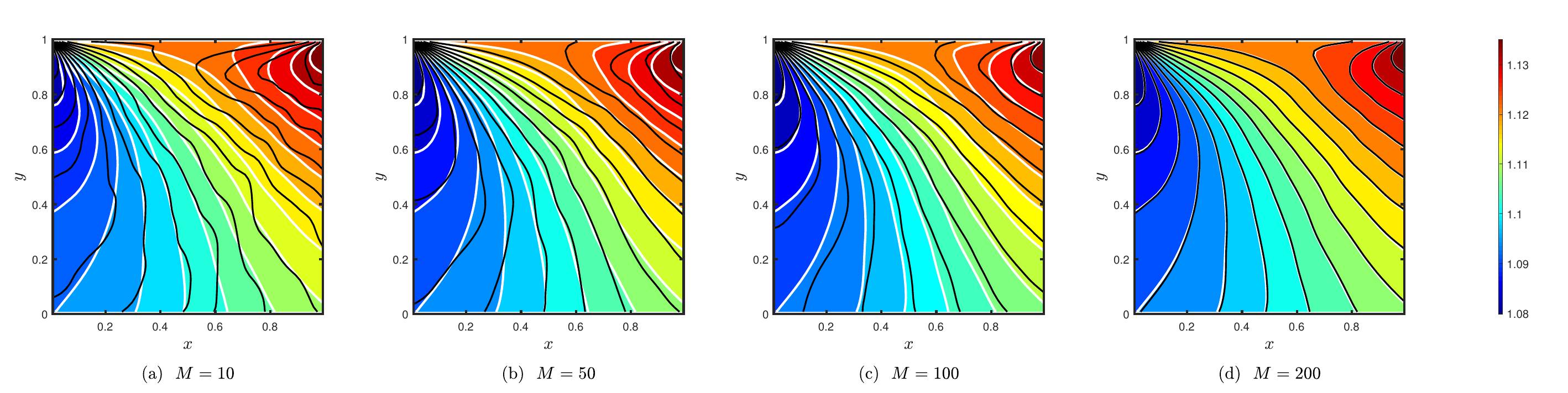}}~~
 \caption{The temperature contour for the $2\mathrm{D}$ lid-driven cavity flow at $\mathrm{Kn} = 10.0$. White solid lines with the colored background: Reference data; black solid lines: SDV-DUGKS results. (a) $M=10$; (b) $M=50$; (c) $M=100$; (d) $M=200$.}
\label{comparison_T_327}
\end{figure}

To capture the flow behavior with non-equilibrium effects, the original DUGKS employs a uniform discretization of the velocity space $[-4\sqrt{2RT_{\mathrm{ref}}}, 4\sqrt{2RT_{\mathrm{ref}}}]^2$ into $81^2$ and $101^2$ points for the $\mathrm{Kn}=1.0$ and $10.0$ cases, respectively, based on the Newton–Cotes quadrature rule.
Regarding the SDV-DUGKS simulations, the same velocity space is sampled using LHS with $1000$ points for both cases. The convergence criterion follows Eq.~\ref{eq:convergence_criterion_MC}, with a maximum iteration limit of 5000 per realization. 
Averages are performed over $M = 100$ and $M = 200$ realizations for the $\mathrm{Kn}=1.0$ and $10.0$ cases, respectively.
For all simulations in this subsection, the reference solution is computed using the original DUGKS with a refined $256 \times 256$ uniform velocity space mesh.
Figs.~\ref{MicrocavityFlow_u_322} and~\ref{MicrocavityFlow_T_323} present the velocity and temperature profiles along the selected lines, respectively, for DUGKS and SDV-DUGKS at different Knudsen numbers, together with the DSMC results~\cite{venugopal2015unified} and reference DUGKS solution for comparison. 
The results of DUGKS and SDV-DUGKS show good agreement with the reference DUGKS solution and the DSMC results.

Additionally, the memory usage and global relative errors of several macroscopic variables are compared for both SDV-DUGKS and DUGKS, as presented in Table~\ref{tab321}.
Using $1000$ sampled discrete velocities per realization, SDV-DUGKS achieves comparable accuracy while requiring only $1/10$ and $1/7$ of the memory usage of DUGKS for cases with $\mathrm{Kn}=10.0$ and $1.0$, respectively. 
These results demonstrate that SDV-DUGKS effectively reduces memory usage compared to DUGKS for subsonic rarefied flows.
Moreover, Figs.~\ref{comparision_314_1} and~\ref{comparision_314_10} present the local relative errors in velocity and temperature along the chosen lines, obtained by averaging over different realizations at $\mathrm{Kn} = 1.0$ and $10.0$, respectively.
A clear overall trend is observed: the local errors decrease as the number of realizations $M$ increases.
The global relative errors in velocity and temperature, obtained by averaging over different numbers of realizations at $\mathrm{Kn} = 1.0$ and $10.0$, are summarized in Tables~\ref{tab322} and~\ref{tab323}, respectively.
To provide a visual assessment, Figs.~\ref{comparison_T_326} and~\ref{comparison_T_327} present temperature contours from SDV-DUGKS with different numbers of realizations $M$ at $\mathrm{Kn} = 1.0$ and $10.0$, respectively, together with the reference DUGKS solution for comparison. These comparisons confirm that averaging over multiple realizations is an effective strategy for reducing statistical noise and obtaining accurate results in subsonic rarefied flow simulations.

\subsection{supersonic flow around a square cylinder}\label{sec33}
\begin{figure}[!ht]
\centering
  \subfloat{\includegraphics[width=0.45\textwidth]{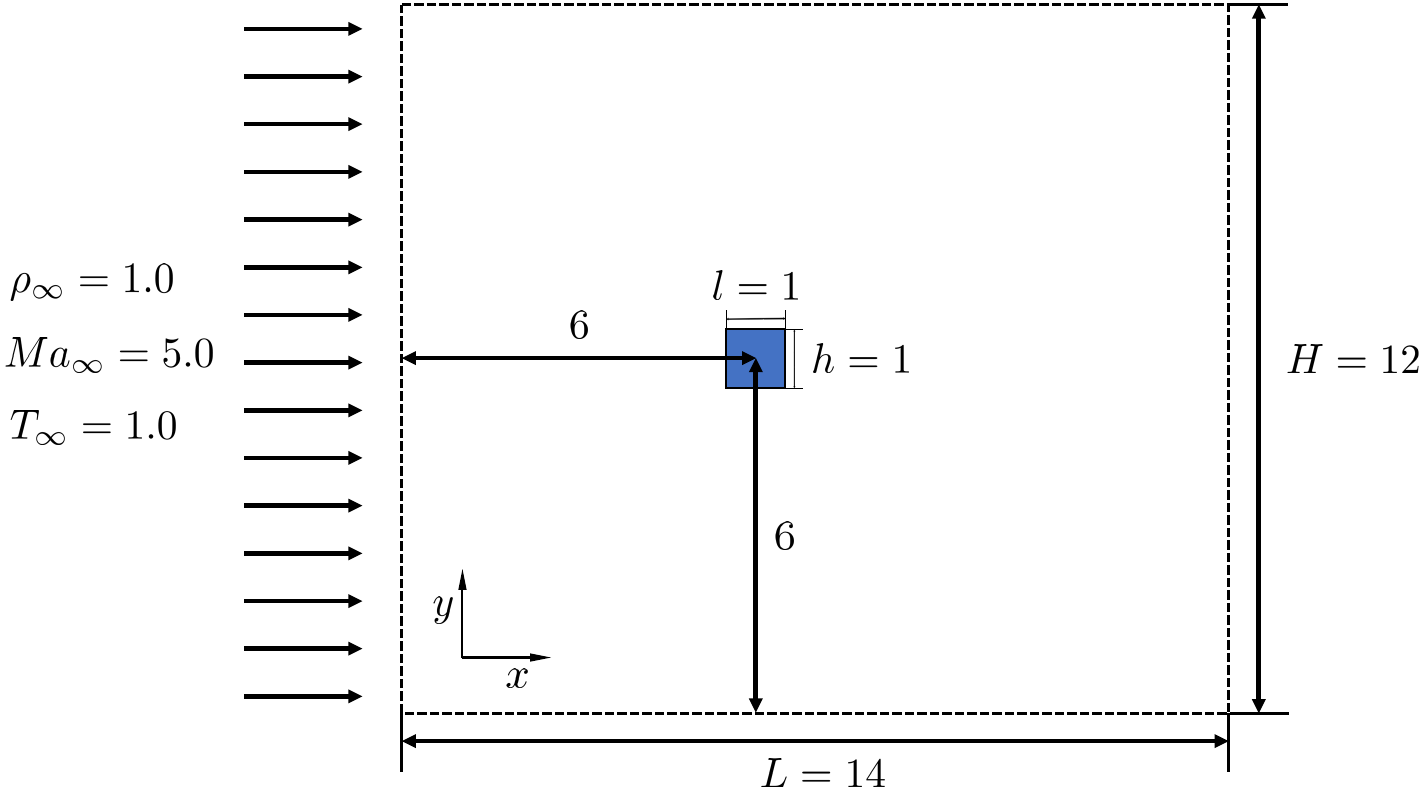}}
 \caption{The schematic of the supersonic flow around a square cylinder.}
\label{Comput_domian_43}
\end{figure}

\begin{figure}[!ht]
\centering
  \subfloat[]{\includegraphics[width=0.45\textwidth]{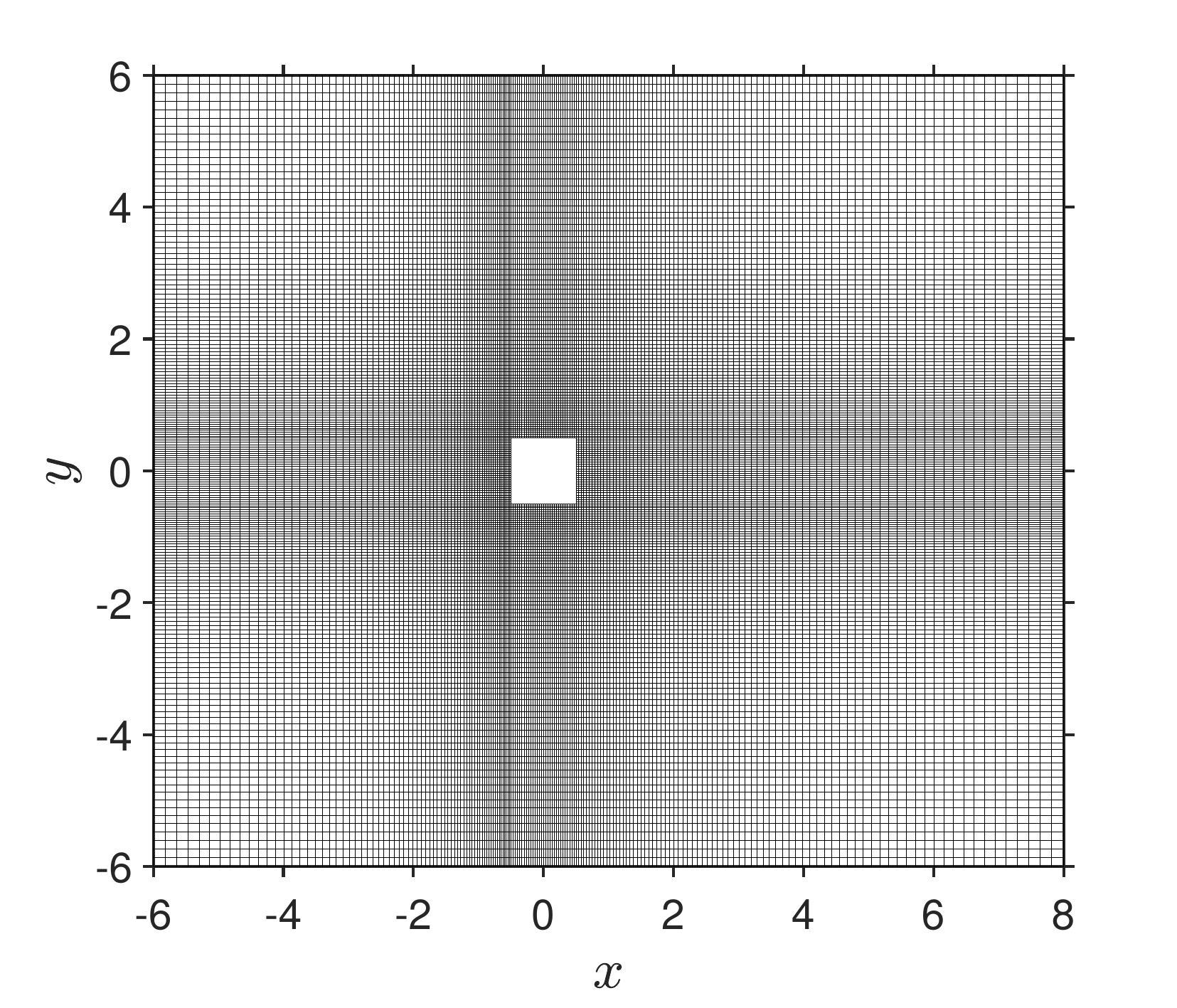}}
    \hspace{8mm}
  \subfloat[]{\includegraphics[width=0.45\textwidth]{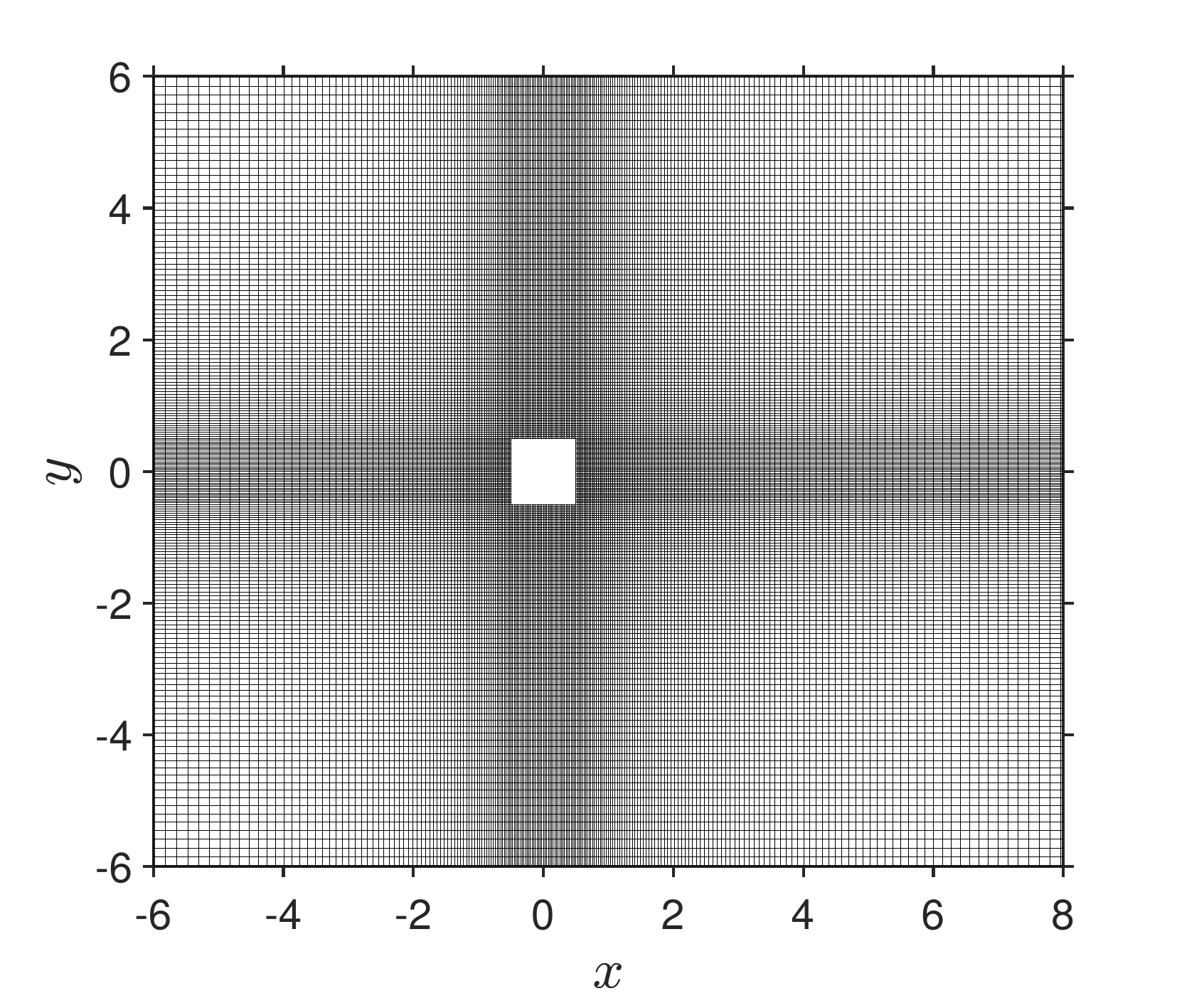}}
 \caption{Meshes of the supersonic flow around a square cylinder. (a) $\text{Kn}=0.1$; (b) \text{Kn}=1.0.}
\label{Mesh_0.1/10_0.01}
\end{figure}

The supersonic flow around a square cylinder is simulated to evaluate the memory reduction performance of SDV-DUGKS at high Mach numbers..
The square cylinder has a side length of $1$, and its wall is maintained at a constant temperature $T_w=1.0$. 
As illustrated in Fig.~\ref{Comput_domian_43}, the computational domain is a $14 \times 12$ rectangular region.
The inlet flow is directed along the positive $x$-direction with a velocity given by $u_\infty=\text{Ma}_\infty\sqrt{\gamma{R}{T_\infty}}$, where the Mach number $\text{Ma}_\infty$ and the specific heat ratio $\gamma$ are $5.0$ and $5/3$, respectively. 
Other macroscopic quantities of the inlet flow, such as density $\rho_\infty =\rho_\text{ref}= 1.0$ and temperature $T_\infty = T_\text{ref}=1.0$, are fixed at these reference values.
The dynamic viscosity is given by $\mu=\mu_{\text{ref}}(T/T_{\text{ref}})^{\omega}$ with $\omega = 0.81$ and $T_\text{ref} = 1.0$. 
Here, the reference viscosity $\mu_{\text{ref}}$ is defined as
in Eq.~\eqref{eq:referenceviscosity}.
Diffuse reflection boundary conditions are applied on all walls of the square cylinder.
To further assess the memory savings of SDV-DUGKS against the original DUGKS, both methods are applied to this problem at Knudsen numbers of $0.1$ and $1.0$.

\begin{figure}[!ht]
\centering
  \subfloat[]{\includegraphics[width=0.45\textwidth]{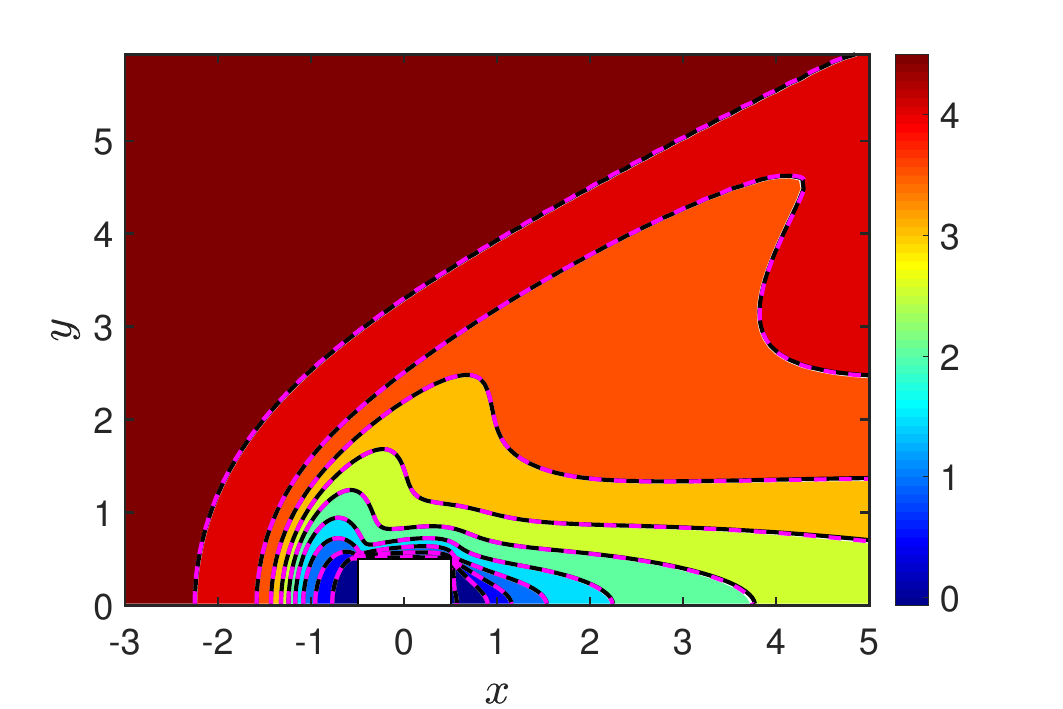}}~~
      \hspace{3mm}
  \subfloat[]{\includegraphics[width=0.45\textwidth]{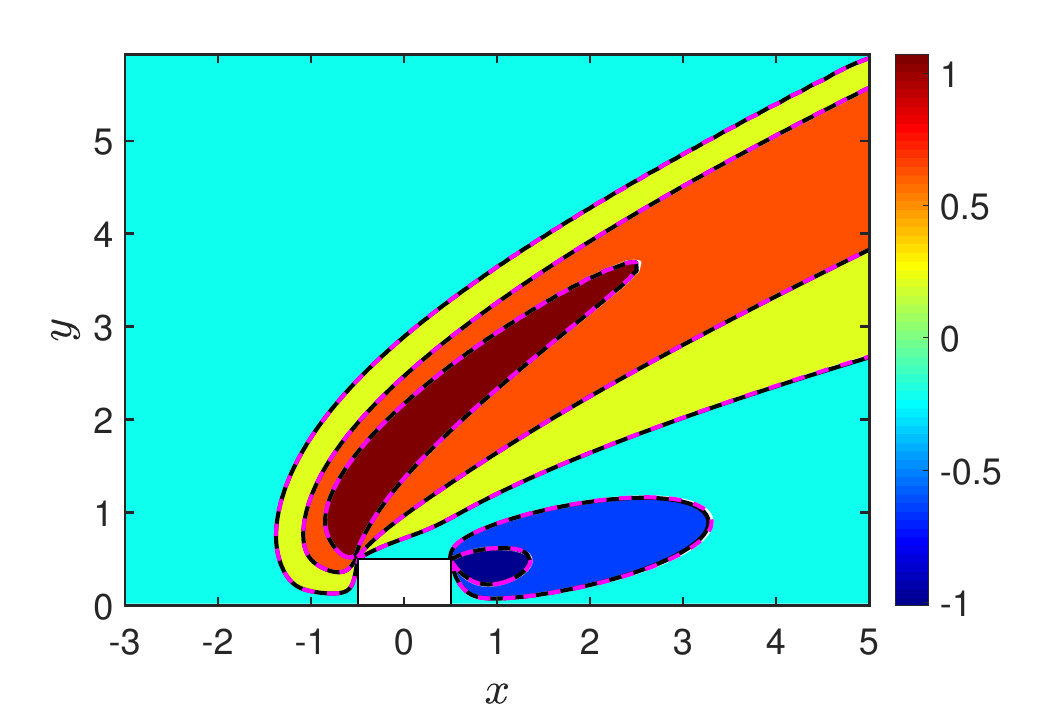}}~~
  \\
  \subfloat[]{\includegraphics[width=0.45\textwidth]{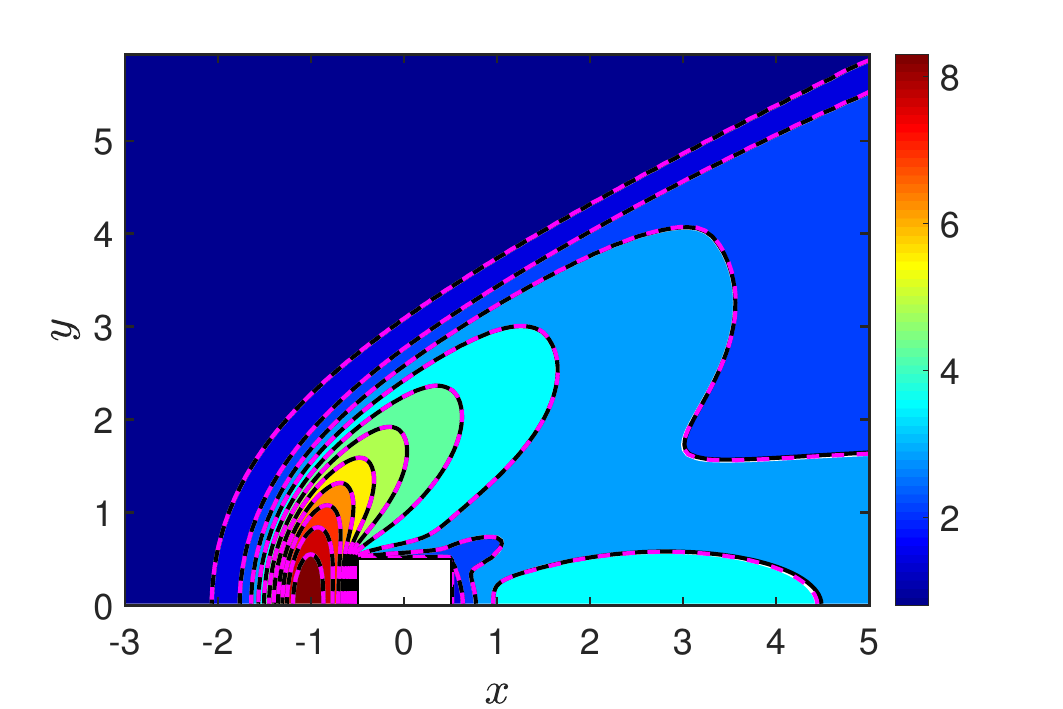}}~~
 \caption{Contours of the $u_x$-velocity (a), $u_y$-velocity (b), and temperature (c) for the supersonic flow around a square cylinder at $\text{Kn}=0.1$ and $\text{Ma}_\infty=5.0$. White solid lines with the colored background: Reference data; magenta dashed lines: DUGKS; black solid lines: SDV-DUGKS.}
\label{Contours_333}
\end{figure}

\begin{figure}[!ht]
\centering
  \subfloat[]{\includegraphics[width=0.55\textwidth]{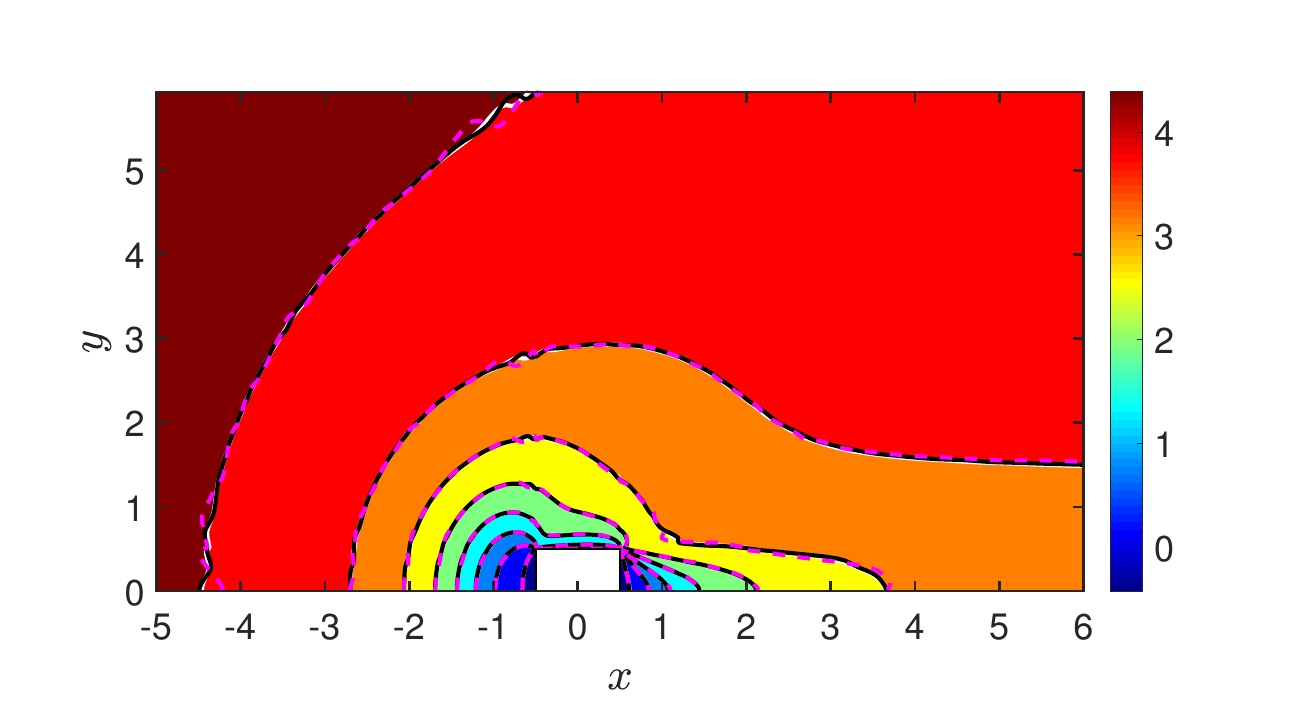}}~~
      \hspace{1mm}
  \subfloat[]{\includegraphics[width=0.55\textwidth]{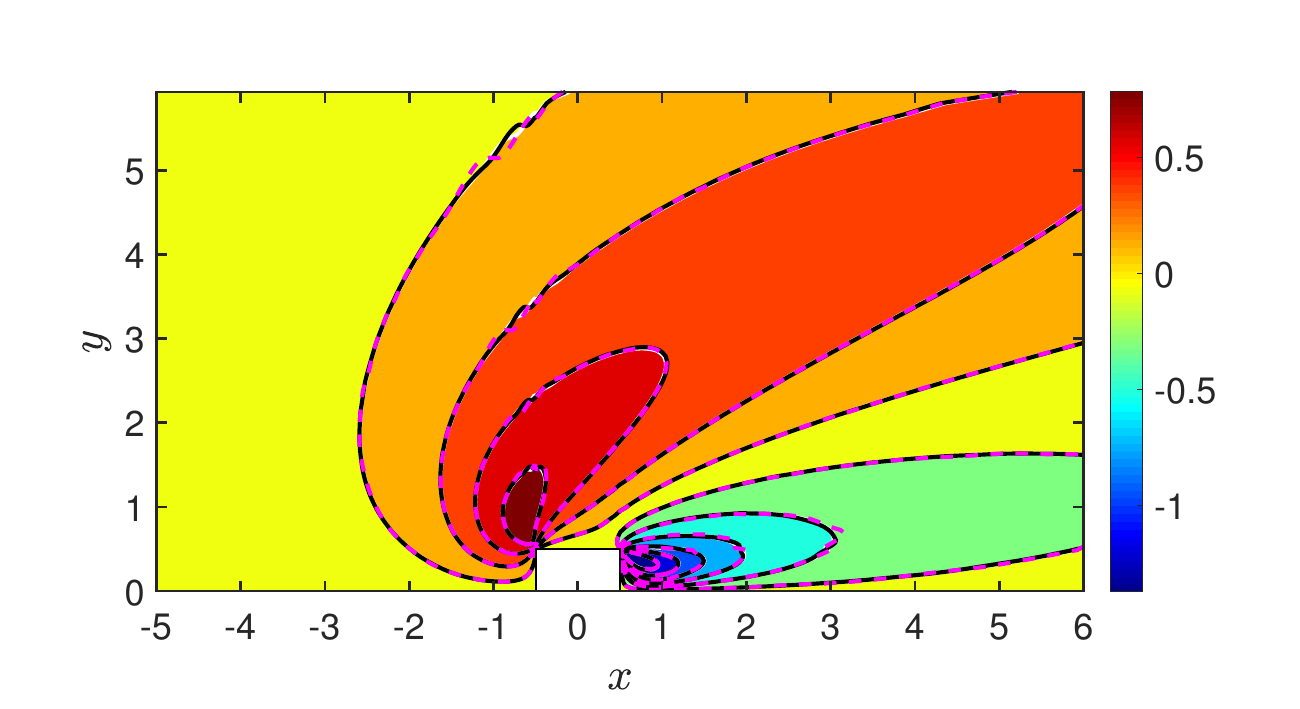}}~~
  \\
  \subfloat[]{\includegraphics[width=0.55\textwidth]{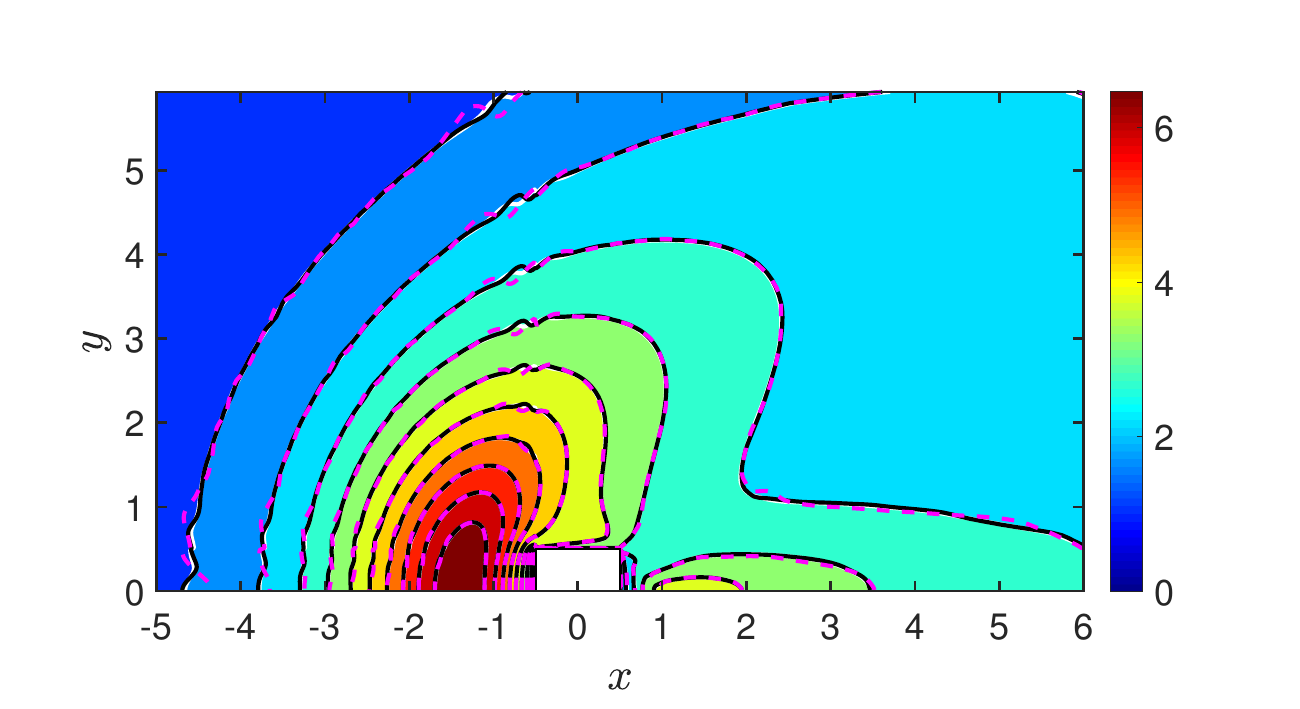}}~~
 \caption{Contours of the $u_x$-velocity (a), $u_y$-velocity (b), and temperature (c) for the supersonic flow around a square cylinder at $\text{Kn}=1.0$ and $\text{Ma}_\infty=5.0$. White solid lines with the colored background: Reference data; magenta dashed lines: DUGKS; black solid lines: SDV-DUGKS.}
\label{Contours_334}
\end{figure}

\begin{table}[!ht]
\caption{Computational cost and relative global error of DUGKS and SDV-DUGKS for supersonic flow around a square cylinder at $\text{Ma}_\infty=5.0$ and various Knudsen numbers.}
\centering
\setlength{\tabcolsep}{2.8mm}
\renewcommand\arraystretch{1.4}
\begin{tabular}{c c c c c c c c c c}
 \toprule
 \multirow{2}{*}{Case} & \multirow{2}{*}{Method} & \multirow{2}{*}{\makecell[c]{Velocity \\ Point}} & \multirow{2}{*}{$M$}
 & \multicolumn{2}{c}{Memory}  & \multicolumn{4}{c}{$E(\%)$} \\
 \cmidrule(lr){5-6} \cmidrule(lr){7-10}
 & & & & Total (GB) & Ratio & $\rho$ & $T$ & $\bm{q}$ & $\bm{\tau}$ \\
 \midrule
 \multirow{2}{*}{$\mathrm{Kn} = 0.1$} & DUGKS & $101^2$ & 1 & 23.870 & 1.0 & 0.116 & 0.092 & 0.122 & 0.127 \\
 \cmidrule(lr){2-10}
 & SDV-DUGKS & 2000 & 50 & 4.745 & $\sim 1/5$ & 0.167 & 0.244 & 0.172 & 0.221 \\
  \midrule
 \multirow{2}{*}{$\mathrm{Kn} = 1.0$} & DUGKS & $101^2$ & 1 & 30.387 & 1.0 & 0.568 & 0.647 & 1.468 & 1.718 \\
 \cmidrule(lr){2-10}
 & SDV-DUGKS & 2000 & 50 & 6.046 & $\sim 1/5$ & 0.596 & 0.591 & 0.831 & 1.192 \\
 \bottomrule
\end{tabular}
\label{tab331}
\end{table}

First, we examine the case at $\text{Kn}=0.1$.
The computational domain is discretized into $33300$ cells. Each edge of the square cylinder is uniformly resolved with $30$ cells, and the grid spacing gradually increases toward the far-field boundaries. Specifically, $70$ cells with a growth ratio of $1.03$ are used along the upstream $x$-axis, while $80$ cells with a growth ratio of $1.02$ are employed along the downstream $x$-axis and the $y$-axis, as illustrated in Fig.~\ref{Mesh_0.1/10_0.01}(a).
In this numerical test, the velocity space domain $[-15,15]\times[-15,15]$ is uniformly discretized into $101^2$ velocity points based on the Newton-Cotes quadrature rule for the original DUGKS. 
The SDV-DUGKS simulation, by contrast, samples the same domain with $2000$ points via LHS. Convergence is declared when the criterion in Eq.~\ref{eq:convergence_criterion_MC} is met, with a maximum iteration limit of $30000$ per realization. Averages are computed over $M = 50$ independent realizations.
Fig.~\ref{Contours_333} presents the contours of the $u_x$-velocity, $u_y$-velocity, and temperature for DUGKS and SDV-DUGKS in the half computational domain at $\text{Kn}=0.1$ and $\text{Ma}_\infty=5.0$, alongside the reference DUGKS solution obtained with a uniform $200^2$ velocity space mesh. 
Both DUGKS and SDV-DUGKS show excellent agreement with the reference solution.

To evaluate the methods at a higher Knudsen number, we conduct a similar comparison at $\text{Kn}=1.0$. 
The computational domain is discretized into $42400$ cells, with each edge of the square cylinder uniformly divided into $40$ grid cells. Grid stretching is applied toward the domain boundaries: $70$ cells (growth ratio $1.03$) on the upstream $x$-axis, $110$ cells (growth ratio $1.02$) on the downstream $x$-axis, and $80$ cells (growth ratio $1.02$) on the $y$-axis.
In this numerical test, the velocity space domain $[-15,15]\times[-15,15]$ is uniformly discretized into $101^2$ velocity points based on the Newton-Cotes quadrature rule. 
For SDV-DUGKS, the same domain is sampled using LHS with $2000$ points, employing the same convergence criterion and a maximum of $30000$ iterations per realization; averages are again computed from $M = 50$ realizations.
Figure~\ref{Contours_334} shows the corresponding contours for DUGKS and SDV-DUGKS in the half computational domain at $\text{Kn}=1.0$ and $\text{Ma}_\infty=5.0$, against the reference DUGKS solution with a uniform $200^2$ velocity space mesh. 
Compared to the original DUGKS results, SDV-DUGKS results not only achieves good agreement with the reference DUGKS solution but also exhibits reduced ray effects.

A detailed comparison of the computational cost and accuracy between DUGKS and SDV-DUGKS across all test cases is summarized in Table~\ref{tab331}. 
For the cases with Knudsen numbers of $0.1$ and $1.0$, SDV-DUGKS achieves comparable accuracy while using only $1/5$ of the memory required. 
These results confirm the effectiveness of SDV-DUGKS in reducing memory usage for supersonic rarefied flow simulations.
However, a limitation is observed: the performance of SDV-DUGKS degrades in supersonic flows compared to low-speed (subsonic) cases. 
We attribute this to the simple sampling strategy employed, which does not adequately resolves the complex velocity distributions characteristic of high-speed flows.
To address this, future work will focus on improving the sampling efficiency of SDV-DUGKS, followed by systematic numerical tests to develop an effective importance sampling strategy.

\section{Conclusion}\label{sec4}
In this work, a memory-efficient deterministic method for multiscale gas flows is proposed, based on an ensemble-of-subproblems strategy with stochastic discrete velocities. 
To be concrete, the proposed method replaces the conventional large deterministic velocity set with multiple small random velocity sets. Each random set defines a subproblem, which is solved by a deterministic multiscale numerical scheme that computes macroscopic moments
via Monte Carlo integration. 
The final flow field is obtained by arithmetic averaging over all subproblems.
This strategy transforms the originally computationally expensive problem into a series of efficiently solvable subproblems.
Consequently, the proposed method can significantly reduce memory requirements for high-Mach-number and strongly non-equilibrium flow simulations.

The performance of SDV-DUGKS is comparatively evaluated against the original DUGKS through three numerical tests: the $1\text{D}$ shock structure, the $2\text{D}$ lid-driven cavity flow, and the $2\text{D}$ supersonic flow past a square cylinder.
The main findings are summarized as follows:
\begin{enumerate}[(1)]
\item The proposed method serves as an effective and accurate tool for simulating multiscale gas flows.
\item In $2\text{D}$ low-speed rarefied flow case, SDV-DUGKS reduces memory usage by approximately $90\%$ compared with the original DUGKS while maintaining comparable accuracy.
\item In $2\text{D}$ supersonic rarefied flow case, SDV-DUGKS achieves a memory reduction of more than $80\%$ relative to the original DUGKS with comparable accuracy.
\end{enumerate}
Therefore, the proposed method exhibits strong potential to mitigate the curse of dimensionality that currently hinders deterministic multiscale numerical schemes from being applied to engineering problems.

This work represents a preliminary attempt. 
Future work will focus on further improving statistical efficiency, e.g., through importance sampling, Bayesian inference, or other techniques.

%%\nolinenumbers
\section*{Acknowledgments}
Zhaoli Guo acknowledges the support provided by the National Natural Science Foundation of China (grant no.12472290).
Weidong Li is grateful to the support provided by National Key Laboratory of Aerospace Physics in Fluids (grant no.KT-APF-2024-004).

%\section*{References}

\bibliographystyle{elsarticle-num-names}
\bibliography{dugks_mc}

\end{document}